\PassOptionsToPackage{prologue,dvipsnames}{xcolor}
\documentclass[sigconf,10pt,nonacm,screen]{acmart}

\usepackage[english]{babel}
\usepackage{blindtext}
\usepackage{epsfig,endnotes}
\usepackage{graphicx}
\usepackage{xspace}
\usepackage[normalem]{ulem}
\usepackage{multirow}
\usepackage{array}
\usepackage{enumitem}
\usepackage{siunitx}

\setlist{nosep}
\usepackage{pifont}
\usepackage{subcaption}
\usepackage{algorithm}
\usepackage{float}
\usepackage[noend]{algpseudocode}
\usepackage{listings}
\usepackage{amsmath} 
\usepackage{hyperref}
\hypersetup{colorlinks={true},linkcolor={violet},citecolor=blue}
\usepackage{cleveref}
\usepackage[font=small]{caption}
\usepackage[font=small]{subcaption}
\usepackage[dvipsnames]{xcolor}
\usepackage{pythonhighlight}
\crefname{section}{}{\S\S}
\crefdefaultlabelformat{\S#2#1#3}
\Crefformat{figure}{#2{}Figure #1{}#3}

\newcommand{\sys}{{{OrchestrRL}}\xspace}

\newcommand{\fabric}{{{RFabric}}\xspace}

\newcommand{\train}{{\texttt{Train}}\xspace}
\newcommand{\gen}{\texttt{Gen}\xspace}

\newcommand{\weightsync}{{Weight Sync}\xspace}

\begin{document}

\date{}

\title{\Large \bf \sys: Dynamic Compute and Network Orchestration for Disaggregated RL}

\author{\large
    Xin Tan$^1$,
    Yicheng Feng$^1$,
    Yu Zhou$^2$,
    Yimin Jiang$^2$,
    Yibo Zhu$^2$,
    Hong Xu$^1$
}
\affiliation{\large
    \institution{$^1$The Chinese University of Hong Kong, $^2$StepFun}
    \city{}
    \country{}
}

\begin{abstract}

Disaggregating the generation and training stages in RL is widely adopted to scale LLM post-training.
There are two critical challenges here. 
First, the generation stage often becomes a bottleneck due to dynamic workload shifts and severe execution imbalances. 
Second, the decoupled stages result in diverse and dynamic network traffic patterns that strain the conventional static fabric.

We build \sys to orchestrate dynamically both compute and network in disaggregated RL. 
\sys employs an adaptive compute scheduler that adjusts parallelism configuration to match changing workload characteristics within and across generation steps. 
\sys adopts a reconfigurable optical-electrical fabric called \fabric : It leverages optical circuit switches to reconfigure the aggregation and core layers of the topology on demand, tailoring bandwidth resources to the unique communication patterns across various phases of training, generation, and weight synchronization.
Evaluated on a 64-H800 GPU testbed, \sys demonstrates up to a 1.42$\times$ throughput improvement over static baselines. 
Using a high-fidelity simulator, we also show that \fabric achieves superior performance-cost efficiency at scale over static Fat-Tree networks.

\end{abstract}

\maketitle
\pagestyle{plain}

\section{Introduction}

Reinforcement Learning (RL) has emerged as a pivotal post-training technique, enabling instruction-following and reasoning capabilities for large language models (LLMs)~\cite{gpt5,cladue4,guo2025deepseekr1}. 
The RL workflow differs from pre-training. For each data batch, it alternates between two stages: sample generation (roll-out) and model training. Generation is memory- and bandwidth-bound since it is autoregressive inference. 
Training is compute-bound: it evaluates those sample responses to compute gradients and update model weights. 

This workflow has two defining characteristics: (1) a strict data dependency that training must wait for generation to complete, and (2) divergent resource requirements across the stages. 
State-of-the-art RL frameworks~\cite{fu2025areal, sheng2025verl, zhong2025streamrl} adopt a disaggregated architecture, using specialized GPU clusters for each stage and enabling asynchronous execution, where generation runs with slightly stale model weights. 
Although fully asynchronous execution has been explored~\cite{fu2025areal,sheng2025laminar}, 
this work focuses on the widely adopted \textit{one-step asynchronous} setting~\cite{sheng2025verl, noukhovitch2024asynchronous,zhong2025streamrl}, which offers a practical balance between algorithmic stability and system throughput.

Despite its benefits, disaggregation {suffers from} two primary inefficiencies that limit performance and scalability. 
First, generation often becomes the end-to-end bottleneck due to rapidly shifting workloads and load imbalance.
Within a single step, the workload evolves from many short generations to a long tail of a small number of lingering responses; across steps, response length distributions can drift as training progresses. 
As a result, an efficient parallelism configuration at the early stage can become suboptimal later, and stragglers can dominate the step makespan (\Cref{sec:characterization}). 
These effects are amplified by unpredictable output lengths and varying KV-cache demands, which cause some GPUs to be underutilized while others are overloaded. Unlike online serving that optimizes metrics such as {Time-to-First-Token (TTFT)}, RL generation is an offline batched workload where the objective is to minimize the \emph{step makespan}---time to complete all samples in a generation step. 
However, existing RL systems largely rely on static or manually tuned parallelism strategies~\cite{zhong2025streamrl, sheng2025verl, fu2025areal}, which are poorly suited to these intra-step and inter-step dynamics.

The second inefficiency lies in a mismatch with the network fabric. Training and generation stages impose diverse and intense traffic patterns. 
Training requires high bisection bandwidth for efficient collective communications (e.g., all-reduce for Data Parallelism (DP), all-to-all for Expert Parallelism (EP) and all-gather/reduce-scatter for Tensor Parallelism (TP)) which alternates within each training iteration. 
Generation clusters, on the other hand, exhibit bipartite communication patterns, such as M2N traffic from {Attention-FFN Disaggregation (AFD)}~\cite{zhu2025megascale, wang2025step} and all-to-all for EP. Disaggregated RL further incurs periodic weight synchronization between training and generation clusters, which is highly bursty. 
Optical Circuit Switching (OCS)~\cite{liao25mixnet, wang2023topoopt, khani2021sipml} has been recently explored to enable dynamic network topologies that potentially overcome the prohibitive cost and poor utilization of a fixed fabric for dynamic traffic (\cref{sec:characterization}). 
Yet, current proposals focus on pre-training and fail to consider the distinct traffic demands at the generation stage, highlighting a critical gap that must be addressed to achieve scalable and efficient network fabric for end-to-end RL workflow. 

In this work, we present \sys, a novel orchestration system for disaggregated RL that treats \emph{reconfiguration} as a first-class principle to jointly optimize compute and network.

For mitigating the compute bottleneck, \sys reconfigures the parallelism strategy in response to workload shifts and imbalance during generation. 
As workload characteristics (e.g., batch size and the generation-length distribution) evolve within and across steps, the optimal operating point on the concurrency-latency trade-off shifts. 
Thus in \sys a proactive planner runs periodically, identifies the best parallelism configuration (e.g., TP, EP, and AFD with different degrees) that minimizes the expected makespan for the remaining requests, and commits the parallelism change to realign deployment with the current workload. Complementing this, a reactive balancer continuously monitors worker load and performs request migrations to reduce stragglers caused by unpredictable output lengths. 

On the networking side, \sys employs \fabric, a hybrid network fabric tailored to the spatiotemporal dynamics of disaggregated RL. 
\fabric partitions training and generation resources into distinct Points-of-Delivery (PoDs) and dynamically reallocates bandwidth based on real-time communication patterns. 
Its hierarchical hybrid optical-electrical design comprises three layers: (i) intra-PoD fabrics using electronic Top-of-Rack switches and OCS-based aggregation, and (ii) an OCS-based core layer interconnecting PoDs. For training PoDs, \fabric configures high-bandwidth topologies to support intensive collective operations that alternate with model layers (e.g., expert parallelism for MoE and context parallelism for Attention) and training phases (forward, backward and gradient synchronization). For generation PoDs, it optimizes for unique bipartite or localized communication patterns. During model synchronization, \fabric temporarily reconfigures the aggregation and core OCS layers to instantiate a two-level, tree-style dissemination overlay from training PoDs to generation PoDs. 
Crucially, the available slack for hiding circuit reconfiguration overhead varies across stages. \fabric adapts its reconfiguration granularity---reconfiguring when slack permits and avoiding the critical path otherwise---to deliver consistently high performance.

To evaluate \sys, we conduct experiments on a 64-GPU NVIDIA H800 testbed and build RLSim, a high-fidelity simulator for large-scale studies. On the testbed, \sys improves end-to-end training throughput by up to 1.42$\times$ over existing schemes. At scale, simulations show that \fabric matches the performance of an ideal non-blocking Fat-tree while improving cost-efficiency by 1.53$\times$--2.06$\times$. Moreover, \fabric outperforms prior optical designs~\cite{wang2023topoopt,liao25mixnet} and achieves a better performance-cost Pareto frontier.

{We make the following contributions.}

\begin{itemize}[leftmargin=3mm]
    \item We systematically characterize the workload dynamics in disaggregated RL, uncovering generation bottlenecks caused by evolving workload shifts and the mismatch between static network and diverse traffic patterns of training and generation stages and their interaction.

    \item We propose \sys, which optimizes compute and network in disaggregated RL via reconfiguration. \sys includes (i) a compute scheduler that minimizes per-step makespan through dynamic parallelism switching and online load rebalancing, and (ii) \fabric, a hierarchical hybrid optical--electrical fabric that enables on-demand topology materialization for stage-specific communication.

     \item We evaluate \sys's compute scheduler on a 64-GPU testbed and evaluate \fabric via large-scale simulation. Results show that \sys improves training throughput and that \fabric achieves higher cost-efficiency than existing network fabrics.

\end{itemize}

\section{Disaggregated RL Workloads}
\label{sec:characterization}

\begin{figure}[t]
  \centering
  \includegraphics[width=0.445\textwidth]{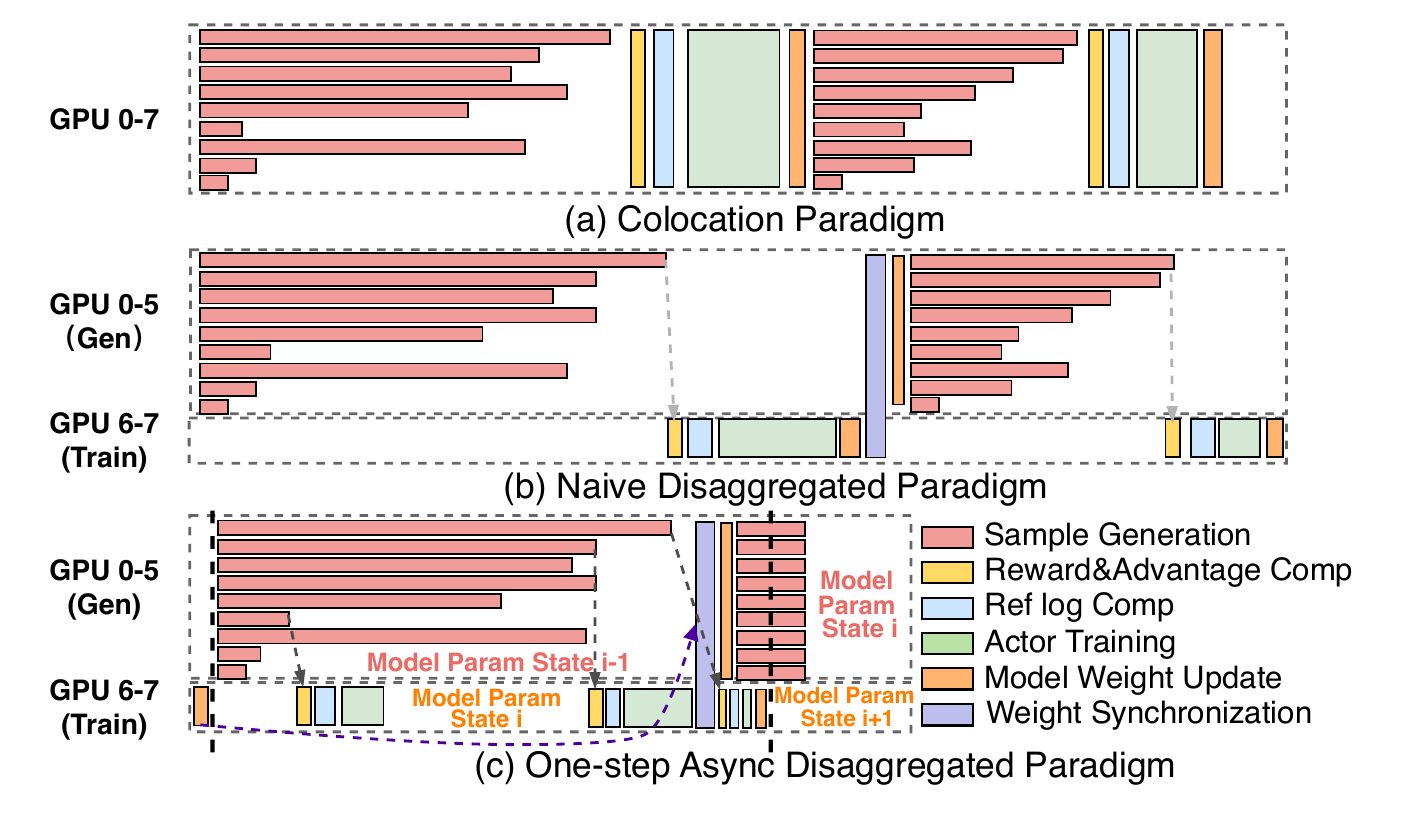}
  \vspace{-0.08in}
\caption{RL workflow in various paradigms with GRPO~\cite{shao2024deepseekmathgrpo}.}
\label{fig:rl_workflow} 
\vspace{-4mm}
\end{figure}

RL systems have evolved from a co-located design to disaggregation. In co-location as in Figure~\ref{fig:rl_workflow}(a)~\cite{sheng2025verl,hu2024openrlhf}, memory-bound generation (\gen) and compute-bound training (\train) share the same GPU bundle, and their mismatched resource demands limit utilization and scalability. Naive disaggregation (Figure~\ref{fig:rl_workflow}(b)) enables specialized, independently scalable clusters, but synchronous execution creates significant pipeline bubbles. One-step asynchronous disaggregation (Figure~\ref{fig:rl_workflow}(c)){~\cite{sheng2025verl, noukhovitch2024asynchronous,zhong2025streamrl}} overlaps clusters on different batches, reducing bubbles and improving throughput, at the cost of using a one-step-stale policy for \gen.

Here we present a detailed characterization of disaggregated RL workloads with {veRL}~\cite{sheng2025verl} under the prevalent one-step off-policy configuration. Unless otherwise noted, we evaluate a Qwen-2.5 14B model with GRPO on the {openr1-math-220k} dataset~\cite{OpenR1-Math-220k} over 72 H800 GPUs: a 32-GPU \train cluster and a 40-GPU \gen cluster. 
\train uses Megatron-LM~\cite{Megatron-LM} with Pipeline Parallelism (PP)=2, DP=2 and TP=8, while the \gen side runs vLLM~\cite{kwon2023vllm} with TP=8 and DP=5.

\subsection{Generation as a Bottleneck}
\label{sec:compute_characterization}

\begin{figure}[t]
  \centering
  \begin{subfigure}[t]{0.38\linewidth}
    \centering
     \includegraphics[width=\textwidth]{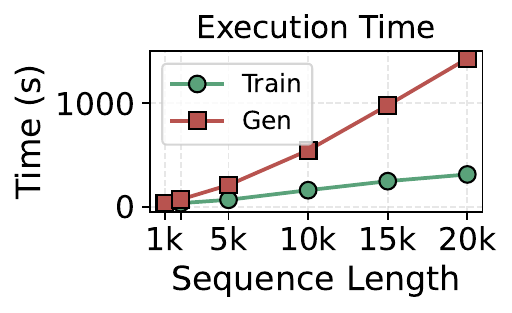}
 \vspace{-0.22in}
 \caption{{Execution times w.r.t.} sequence length; batch size 512.}
 \label{fig:compute_diff}
  \end{subfigure}
  \hfill
  \hspace{-0.25in}
  \begin{subfigure}[t]{0.59\linewidth}
    \centering
    \includegraphics[width=\linewidth]{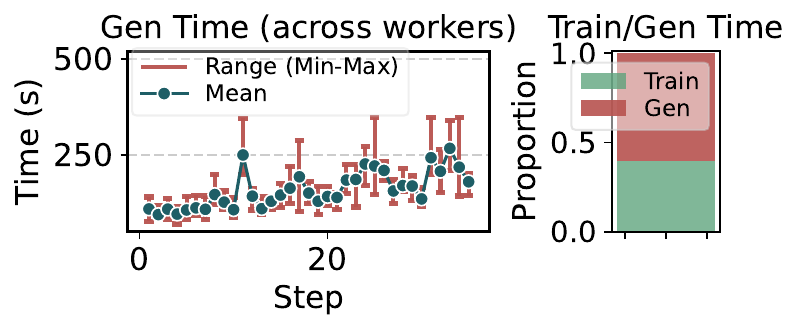}
     \vspace{-0.22in}
    \caption{Gen worker time profiles. The long tail of samples increases skew.}
    \label{fig:gen_worker_time}
  \end{subfigure}
   \vspace{-0.15in}
  \caption{Effect of sequence length on performance. }
  \label{fig:gen_overview}
\end{figure}

\noindent\textbf{\gen on the critical path.} The \gen stage is far more sensitive to increase in sequence length than \train. As shown in Figure~\ref{fig:compute_diff}, training and generation have distinct performance profiles. \train, which involves full forward and backward passes on large batches, is predominantly compute-bound, making it well-suited for modern accelerators. In contrast, \gen is memory bandwidth-constrained and depends on an iterative decoding process, heavily impacted by sequence length. Consequently, \gen frequently dominates the step makespan, taking 1.49$\times$ longer than \train in our workload (Figure~\ref{fig:gen_worker_time}) on average. This gap can persist even with additional \gen resources, because \gen acts as the producer and its cost scales directly with sequence length.

\begin{figure}[t]
  \centering
  
\begin{subfigure}[t]{0.565\linewidth}
    \centering
\includegraphics[width=\textwidth]{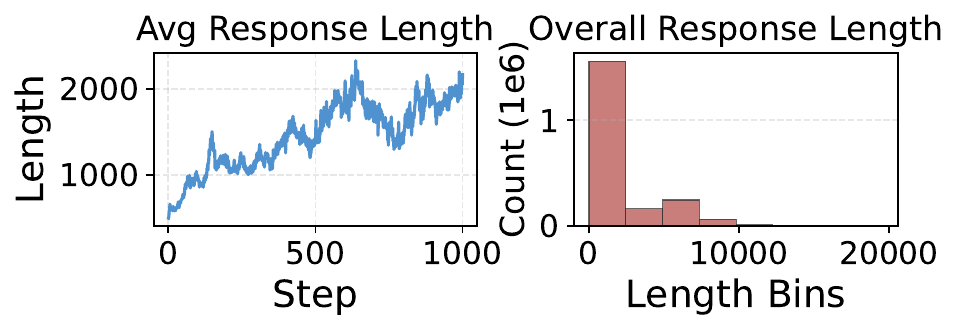}
 \vspace{-0.25in}
 \caption{Distribution of response lengths with a heavy tail.}
 \label{fig:response_length_dist}
  \end{subfigure}
   \hfill
  \hspace{-0.5in}
  \begin{subfigure}[t]{0.385\linewidth}
    \centering
  \includegraphics[width=\linewidth]{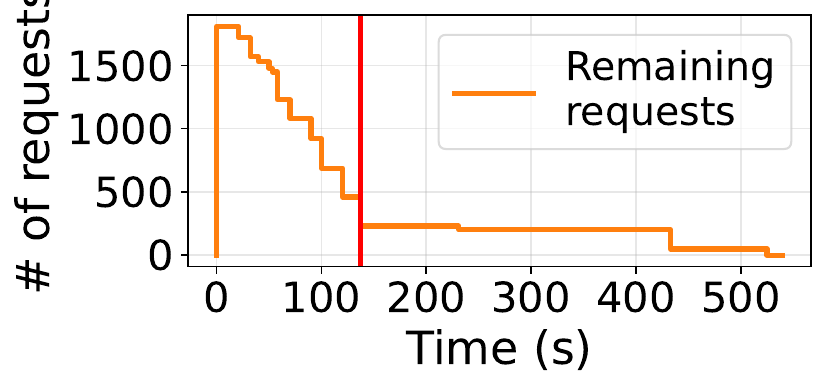}
   \vspace{-0.25in}
    \caption{Remaining request number in a \gen{} step.}
    \label{fig:remaining_requests}
  \end{subfigure}
   \vspace{-0.1in}
  \caption{Generation dynamics shaped by request completion and length variability. %
  }
  \label{fig:seq_len}
\end{figure}

\noindent\textbf{Root causes of inefficiency.} (1) \uline{Heavy-tailed and evolving response length distribution}, as shown in Figure~\ref{fig:response_length_dist}. Most responses are of moderate length (under 7.5K tokens), but a small fraction of stragglers require substantially more decoding steps and can delay overall processing (Figure~\ref{fig:remaining_requests}), consistent with observations in~\cite{zhong2025streamrl,zhong2024rlhfuse}. 
(2) \uline{Fixed parallelism for
changing workloads.} Static configurations cannot adapt to shifts in batch size or response length, leading to suboptimal performance. Even
within a single generation step, the best parallelism configuration can change across phases under a fixed 8-GPU budget
(with \(\text{TP}\times\text{DP}\) constant and batch size 512). We select two snapshots within the step: in the early
phase, the step completes fastest with TP=2 (79.2s, vs.\ 96.3s/138.7s for TP=4/8), while in the tail phase it is fastest
with TP=8 (165.1s, vs.\ 188.3s/208.4s for TP=4/2). Clearly, any single fixed configuration is inefficient across the step.
(3) \uline{Imbalance across workers due to response length variability}. As shown in Figure~\ref{fig:gen_worker_time}, workers handling shorter responses finish earlier, while those processing longer responses become stragglers (with the max-to-min average time ratio reaching 1.58$\times$ in the sampled workload). Additionally, KV cache utilization is also skewed across workers.

\subsection{Asymmetric and Bimodal Traffic }
\label{sec:communication_characterization}

\begin{figure}[t]
 \centering
 \includegraphics[width=0.405\textwidth]{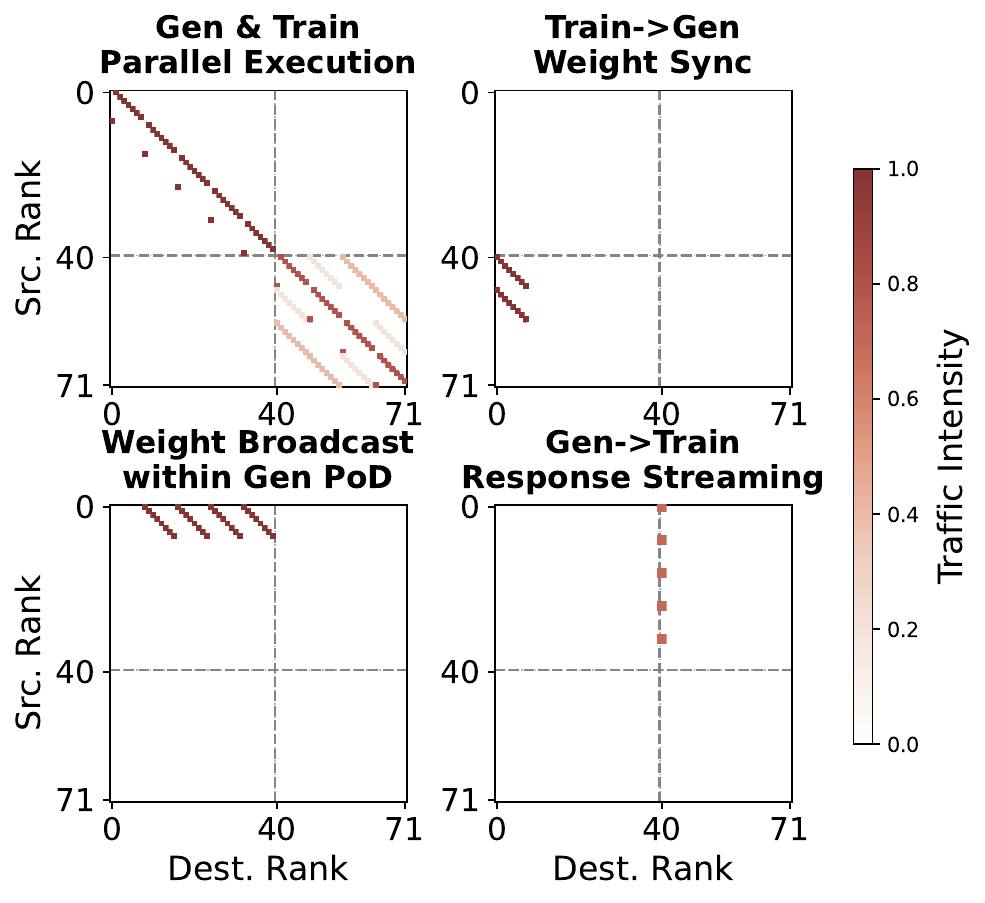}
 \vspace{-0.15in}
 \caption{Spatial network traffic across RL stages. For weight synchronization, we use an optimized two-stage scheme: the \train DP group transmits weights once to the DP-0 group of each \gen pod, after which each \gen DP-0 broadcasts locally to its DP peers. 
 }
 \label{fig:network_spatial_temporal}
 \vspace{-4mm}
\end{figure}

\noindent\textbf{Spatial heterogeneity.}
Communication is highly asymmetric across stages during concurrent \gen and \train execution (Figure~\ref{fig:network_spatial_temporal}). The \train cluster (Ranks 40--71) exhibits structured collective patterns spanning multiple parallelism dimensions: frequent TP all-reduces within model-parallel groups, DP gradient synchronization across DP replicas, and PP send/recv traffic between adjacent pipeline stages (and, if in MoE (Mixture-of-Experts) settings, additional EP all-to-all exchanges). In contrast, the \gen cluster (Ranks 0--39) communicates almost entirely within small, disjoint TP groups, with negligible traffic across groups. Cross-cluster traffic is sparse and bursty, occurring mainly at stage boundaries for weight synchronization, plus small messages from \gen to \train to stream generated samples back for training. 

\begin{figure}[t]
 \centering
\includegraphics[width=0.445\textwidth]{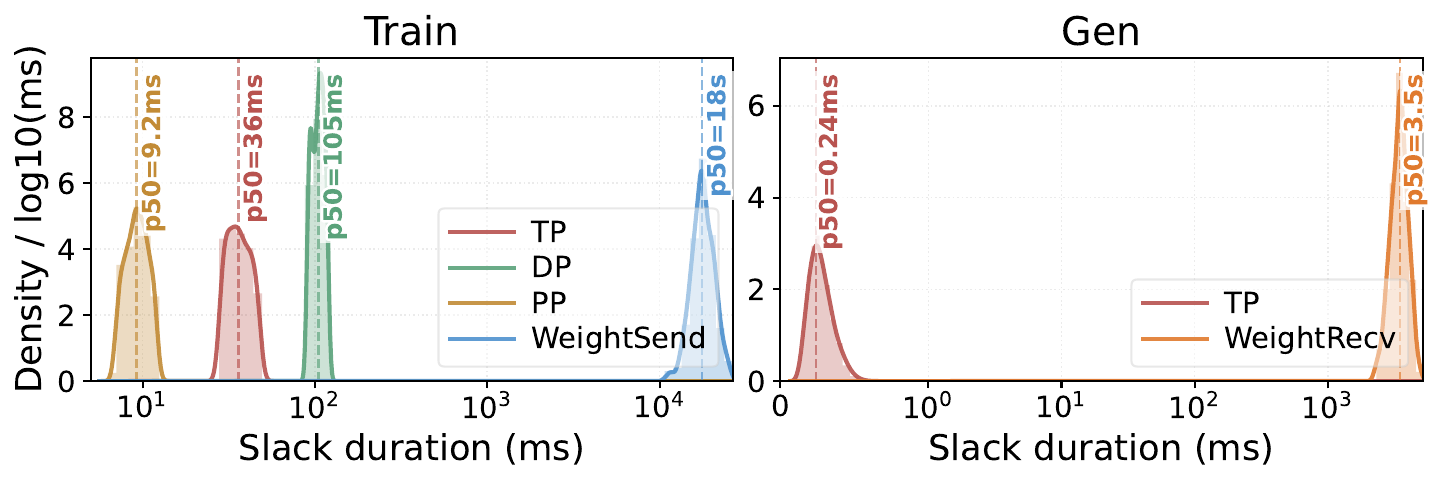}
   \vspace{-0.08in}
 \caption{Distribution of slack durations across communication operation types.}
 \label{fig:network_window}
 \vspace{-3mm}
\end{figure}

\noindent\textbf{Temporal heterogeneity and reconfiguration slack.}
Communication timing and intensity vary across stages, creating intermittent \emph{slack} windows for OCS reconfiguration. We define \emph{communication slack} for a collective event $e_k$ as $\Delta t_k = t_{\text{start}}(e_k) - t_{\text{end}}(e_{k-1})$, the time since the previous communication event. Figure~\ref{fig:network_window} shows that slack is highly heterogeneous with a bimodal distribution.

One mode corresponds to dense phases dominated by model-parallel collectives, where slack remains consistently small, leaving little room for disruption. For example, during \gen, TP collectives occur almost back-to-back, with a median slack below 0.5ms, owing to small inputs and short compute intervals. During \train, TP/PP/DP collectives are still frequent, but longer compute kernels open up larger slack windows (typically tens to hundreds of milliseconds), allowing reconfiguration to be scheduled more judiciously. The other mode corresponds to sparse weight synchronization (\texttt{WeightSend}/\texttt{WeightRecv}), where events are separated by long idle periods (seconds or more), providing ample opportunities for reconfiguration.

\subsection{Design Opportunities}
\label{subsec:oppo}

Our analysis highlights two defining characteristics of disaggregated RL. First, the \gen stage (the workflow producer) is highly sensitive to response-length variability, leading to dynamic workload shifts and intra-cluster load imbalance. Second, network demand varies substantially across \train, \gen, and weight synchronization, and changes over time.

\noindent\textbf{Opportunity 1: Adaptive compute scheduling.} 
To adapt to evolving workloads, it is essential to periodically reevaluate parallelism strategies and resource placement for longer-term changes (e.g., batch size and generation length distributions).
Further, dynamic request migration 
can correct short-term imbalances with low overhead.
Together, these two operate at different timescales to sustain high utilization.

Moreover, unlike online serving that prioritizes latency (i.e. TTFT) through the prefill-decode trade-offs, generation in RL reasoning is makespan-oriented. In this context, decoding dominates the critical path during long rollouts, shifting the optimization focus to decoding-centric parallelism with less emphasis on prefill-decoding disaggregation~\cite{zhong2024distserve}.

\noindent\textbf{Opportunity 2: Workload-aware network reconfiguration.}
As shown earlier, disaggregated RL exhibits stage- and parallelism-dependent communication patterns, and our adaptive parallelism (Opportunity~1) further changes collective membership and shifts hotspots over time.
This creates a dilemma: \train benefits from high-bandwidth circuits for bulk-synchronous collectives and periodic \train-\gen synchronization, whereas \gen demands sub-millisecond, always-on connectivity for latency-critical fine-grained traffic.

A monolithic fabric struggles to satisfy both requirements.
An optical network can provide high and reconfigurable bandwidth. 
While commodity OCS devices can switch in tens of milliseconds (Table~\ref{tab:ocs_reconfig_time}), the \emph{end-to-end} time before new links are usable is often dominated by system effects (e.g., clock/data recovery, control-plane actions, and routing updates). For example, MixNet~\cite{liao25mixnet} reports OCS control latencies of tens of milliseconds, yet observes multi-second overheads due to host-side device initialization when the NIC is \emph{directly connected} to the OCS.
If such reconfiguration lies on \gen's critical path, it would dramatically inflate overall latency and degrade performance.
Conversely, relying solely on an electrical network forces the fabric to provision for worst-case global traffic, which is costly and still prone to congestion under bursty demand.

These observations motivate \fabric: a hybrid EPS$-$OCS design that keeps an always-on, host-facing EPS substrate for latency-sensitive traffic, while using an upper-tier OCS to opportunistically provision stage-specific high-bandwidth circuits when slack permits.
By aligning circuit schedules with each stage's dominant communication (e.g., \train forward/backward collectives vs.\ gradient synchronization), \fabric improves utilization and end-to-end performance under a fixed hardware budget.
Moreover, because slack varies widely across stages, \fabric adapts its reconfiguration frequency accordingly.

\section{\sys Overview}

\begin{figure}[t]
 \centering
 \includegraphics[width=0.365\textwidth]{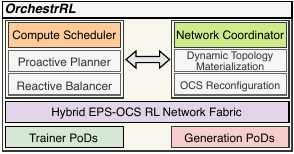}
   \vspace{-0.15in}
\caption{Overview of \sys.}
\label{fig:sys_overview}
\end{figure}

\sys dynamically orchestrates both compute and network resources for disaggregated RL. As shown in Figure~\ref{fig:sys_overview}, \sys combines a co-designed compute--network architecture with a unified control plane.

\noindent\textbf{Disaggregated PoDs, reconfigurable fabric.}
\sys adopts a disaggregated design that partitions training and generation resources into distinct PoDs, enabling specialization and independent scaling. On top of this architecture, \sys deploys a reconfigurable hybrid optical--electrical fabric (\Cref{subsec:fabric}) that adapts its topology to the stage-dependent communication demands of RL.

\noindent\textbf{Dynamic orchestration.}
The unified control plane jointly coordinates compute and network to reduce bottlenecks and improve efficiency. \textit{On the compute side}, an adaptive scheduler tracks \gen workloads, adjusts parallelism configurations, and mitigates stragglers via lightweight request migration across replicas to reduce the \gen step makespan (\Cref{sec:compute_orchestration}). \textit{On the network side}, a workload-aware coordinator translates \emph{workflow intents} (across \train, \gen, and weight synchronization) into topology and bandwidth decisions and reconfigures the fabric on demand (\Cref{sec:network}).

\section{Dynamic Compute Orchestration}\label{sec:compute_orchestration}

This section describes \sys's adaptive compute orchestration. To handle the workload shifts and intra-cluster imbalance identified in \Cref{sec:characterization}, \sys employs a two-level mechanism as summarized in Algorithm~\ref{alg:adaptive_orchestration}.

\begin{itemize}[leftmargin=3mm]
    \item \textbf{Coarse-grained proactive planning} (\Cref{subsec:proactive_planning}): In each \gen step, \sys periodically forecasts the remaining workload, computes the best parallelism configuration that minimizes the makespan, and updates the deployment if beneficial, in order to adapt to the evolving workload shifts. 
    
    \item \textbf{Fine-grained reactive balancing} (\Cref{subsec:reactive_balancing}): On a shorter timescale, \sys also continuously migrates requests away from stragglers to mitigate the transient imbalances caused by unpredictable workload variations.
\end{itemize}

\subsection{Proactive Planning}\label{subsec:proactive_planning}

\begin{algorithm}[t]
\caption{Orchestration for \gen}
\footnotesize
\label{alg:adaptive_orchestration}
\begin{algorithmic}[1]

\State \textbf{Event 1: Periodic proactive planning within each \gen step:}
\Statex \textit{{\color{blue}{\# Global planning for slow-evolving workload shifts.}}}
\State $\mathcal{R} \leftarrow cluster.GetPendingReqs()$; $\mathcal{C}_{cur} \leftarrow cluster.GetConfig()$
\State $Obj_{cur} \leftarrow \text{EstMakespan}(\mathcal{R}, \mathcal{C}_{cur})$
\State $\{\mathcal{C}_{new}, Obj_{new}, cost\} \leftarrow \text{SolveOptim}(\mathcal{R}, \mathcal{C}_{cur})$ \Comment{Solve Eq.~\ref{eq:objective}}
\If {$Obj_{new} + cost < Obj_{cur} - \epsilon$}\Comment{Gain is sufficient}
    \State $cluster.Reconfigure(\mathcal{C}_{new})$
\EndIf

\Statex

\State \textbf{Event 2: On each monitoring tick between proactive cycles:}
\Statex \textit{{\color{blue}{\# Handle transient imbalance under the current configuration.}}}
\State $\mathcal{L} \leftarrow cluster.GetLoadIndices()$
\State $\mathcal{R} \leftarrow cluster.GetPendingReqs()$
\If {$\text{Imbalance}(\mathcal{L}) > \theta$}
    \State $\mathcal{M} \leftarrow \text{IdentifyMigrationCandidates}(\mathcal{L}, \mathcal{R})$
    
    \State $cluster.ExecBalance(\mathcal{M})$ \Comment{Request-state migration}
\EndIf

\end{algorithmic}
\end{algorithm}

During each \gen step, we monitor request progress and assess whether reconfiguring the parallelism can reduce the expected completion time of pending requests. Rather than predicting per-request output lengths---which is both impractical and challenging~\cite{zhong2025streamrl,qin2025seer}---we maintain an online estimate
of the \emph{step-level} remaining-workload histogram.
At the beginning of each step, we forecast this histogram using an ARIMA model~\cite{box2015time} trained on output-length statistics from recent steps.
As requests are completed, this distribution is dynamically updated by removing finished requests.
With the updated distribution and per-request generation progress, we estimate the remaining decoding workload. Accounting for KV-cache constraints, we select the parallelism configuration that minimizes the makespan of the remaining work, and apply it only when the expected gains outweigh reconfiguration overheads.

\noindent\textbf{Problem formulation.}
Let $\mathcal{R}$ be the set of pending requests and $\mathcal{K}$ the set of \emph{logical instances}.
Each instance $k\in\mathcal{K}$ represents a schedulable deployment option: a feasible GPU group together with a parallelism configuration
$y_k\in\mathcal{Y}$ (e.g., TP/EP/AFD and degree), which uses $G(y_k)$ GPUs.
We decide a request-to-instance assignment $\mathbf{X}=[x_{ik}]\in\{0,1\}^{|\mathcal{R}|\times|\mathcal{K}|}$ and the per-instance configuration vector
$\mathbf{Y}=[y_k]$.
For instance $k$, $\mathcal{R}_k(\mathbf{X})=\{i\in\mathcal{R}: x_{ik}=1\}$ denotes the requests assigned to it.
We denote the current setting as $(\hat{\mathbf{X}},\hat{\mathbf{Y}})$.

\begin{align}
   \min_{\mathbf{X}, \mathbf{Y}} \quad &
\max_{k\in\mathcal{K}} \Big(
    T_k\!\left(\mathcal{R}_k,\, y_k\right)
    + C^{\text{mig}}_{k}\!\left(\mathbf{X}, \hat{\mathbf{X}}\right)
    + C^{\text{sw}}_{k}\!\left(y_k, \hat{y}_k\right)
\Big)
\label{eq:objective} \\
    \text{s.t.} \quad 
    & \left| \frac{\sum_{i \in \mathcal{R}} x_{ik} \cdot D^{\text{kv}}(i)}{\text{Cap}^{\text{kv}}(y_k)} - \mu \right| \le \delta, \quad \forall k \in \mathcal{K}, \label{eq:mem_balance} \\
    & \sum_{k \in \mathcal{K}} {G}(y_k) \le G_{\text{total}}, \label{eq:res_cons} \\
    & \sum_{k \in \mathcal{K}} x_{ik} = 1, \quad \forall i \in \mathcal{R}. \label{eq:valid_cons}
\end{align}
The objective~\eqref{eq:objective} minimizes the predicted makespan across instances, while accounting for state-migration cost $C^{\text{mig}}$ for transferring KV-cache and parallel-mode switching cost $C^{\text{sw}}$. 
$T_k(\mathcal{R}_k,y_k)$ estimates the remaining completion time of instance $k$ under configuration $y_k$, capturing batching effects and KV-cache dynamics. 
Constraint~\eqref{eq:mem_balance} balances KV-cache load \emph{normalized by capacity} by keeping each instance's KV utilization,
$\sum_i x_{ik} D^{\text{kv}}(i) / \text{Cap}^{\text{kv}}(y_k)$, within $\mu \pm \delta$; transient overflows are spilled to CPU memory.
Constraint~\eqref{eq:res_cons} enforces the total GPU budget $G_{\text{total}}$, and~\eqref{eq:valid_cons} assigns each request to exactly one instance.

\noindent\textbf{Reconfiguration and state migration costs.} 
Before discussing how to solve this optimization~\eqref{eq:objective}, we ought to take a close look at the overheads of reconfiguring parallelism,  $C^{\text{sw}}$ and $C^{\text{mig}}$.
Executing a new plan takes two steps: updating instance placement and migrating request state. 
First, model weights need to be re-sharded (and loaded if necessary) according to the new parallel layout. 
We estimate the this switching cost $C^{\text{sw}}$ from the amount of weights to redistribute/reload and the offline-profiled bandwidth/latency of the transfer path. 
When supported, we also use GPU virtual memory to pre-map (warm up) weight and KV-cache regions for standby configurations to reduce switch-over latency~\cite{vllm_sleep_mode}, and include this overhead in $C^{\text{sw}}$. 
Next, KV cache migration to the new sharding scheme is performed via either (1) direct transfer using intra-node NVLink/PCIe and inter-node RDMA, or (2) recomputation by replaying the forward pass from cached token IDs. The migration cost $C^{\text{mig}}$ is estimated by comparing KV transfer time with recomputation time, selecting the cheaper option based on an offline-profiled cost model (see Appendix~\ref{app:kv-migration-cost} for details).

\noindent\textbf{Solving the optimization.}
The optimization program~\eqref{eq:objective} involves a search space of combinations of parallel modes and instance groups which also grow with the number of GPUs and pending requests and is expensive to solve exactly. 
We apply three domain-specific heuristics to {reduce the search space and make it practically solvable.}

\textit{First}, we implement the remaining-workload histogram using fixed-size bins (e.g. 256 tokens), collapsing per-request variables into per-bin
aggregates to reduce the optimization dimensionality. We use each bin's upper bound as a conservative length estimate. The histogram predictor forecasts the
per-bin request counts and achieves <9.8\% relative error on our workload.

\textit{Second}, we restrict the candidate configurations to a small, practical set. We prioritize intra-node TP and consider cross-node TP only in whole-node increments (e.g., 8 GPUs). AFD candidates are also pre-selected via offline profiling to retain only a few high-performing Attention-to-FFN ratios.

\textit{Third}, we adapt the candidate set within each generation step as the workload evolves. When the active batch is large and remaining lengths are moderate, we bias toward throughput-optimized modes. As the batch shrinks and long-tail sequences dominate, we prioritize latency-optimized modes to reduce tail completion time. For example, AFD can be effective with large batches by overlapping network communication with compute, but it often underperforms in the long tail scenario, where communication becomes difficult to hide and turns into the critical path.

\subsection{Reactive Balancing}
\label{subsec:reactive_balancing}

Proactive planning addresses structural shifts but struggles with transient imbalances caused by unpredictable output lengths and KV-cache pressure.
To mitigate these transient stragglers, \sys uses a lightweight \textit{Reactive Balancer} that periodically ranks workers by congestion and selectively migrates requests.

Without an oracle for each request's response length, we define a congestion score $\mathrm{LoadIndex}_w$ (higher means more congested) for each \gen worker that combines queue pressure, KV headroom, and observed decoding rate:

\begin{align}
\mathrm{LoadIndex}_w \;=\;
{\frac{|\mathcal{Q}^{run}_w| + |\mathcal{Q}^{wait}_w|}{B^{cap}\!\left(M^{free}_w\right)}}
\;\times\;
{\frac{1}{\widetilde{R}_w}}.
\end{align}
Here, $|\mathcal{Q}^{run}_w|$ and $|\mathcal{Q}^{wait}_w|$ denote the numbers of running and waiting requests. $B^{cap}(M^{free}_w)$ represents the maximum concurrency allowed by the runtime KV allocator based on the current memory margin ($M^{free}_w$). $\widetilde{R}_w$ is the decoding throughput (tokens/s). 

This design is driven by three insights:
(1) \emph{Queue depth is only meaningful relative to KV-limited concurrency.}
Because generation lengths vary across requests, the same queue depth can indicate very different congestion under different KV headroom; we therefore normalize it by $B^{cap}(M^{free}_w)$.
(2) \emph{Service rate matters.}
Even with similar normalized backlogs, workers with different decoding rates can have very different completion times; we include $1/\widetilde{R}_w$.
(3) \emph{Migrations must be KV-feasible.}
A migration is beneficial only if the destination can accommodate the request's KV footprint without reducing the destination worker's stable concurrency (i.e., effective batch size); $\mathrm{LoadIndex}$ ranks candidates, while KV-cache constraints gate feasibility.

To avoid thrashing, we migrate conservatively. Every interval, we compute
$\Delta=\max_w \mathrm{LoadIndex}_w-\min_w \mathrm{LoadIndex}_w$ and trigger migration only if $\Delta>\theta$.
We move requests from the most congested worker to the least congested, prioritizing the waiting queue under a short-context-first policy.
A migration is performed only if the destination has sufficient KV headroom to host the request without reducing its stable batch size.
The procedure stops once $\Delta\le\theta$ or no further requests are feasible under KV constraints.

\begin{table}[t]
\centering
\resizebox{\columnwidth}{!}{%
\begin{tabular}{c|c|ccc|cc}
\hline
\multicolumn{1}{l|}{} & \multirow{2}{*}{\textbf{Ops}} & \multicolumn{3}{c|}{\textbf{Profile}} & \multicolumn{2}{c}{\textbf{Fabrics}} \\ \cline{3-7} 
\multicolumn{1}{l|}{} &  & \textbf{Volume} & \textbf{Frequency} & \textbf{Primitives} & \textbf{\begin{tabular}[c]{@{}c@{}}Possible \\ Domain\end{tabular}} & \textbf{Reconfigurability} \\ \hline
\multirow{5}{*}{\textbf{\begin{tabular}[c]{@{}c@{}}Train.\\ Stage\end{tabular}}} & DP & High & Low & AllReduce & ToR-Agg-Core & High \\
 & TP & Medium & High & AllReduce & HBD/ToR & Medium \\
 & PP & Low & Low & P2P & ToR-Agg & High \\
 & CP & Medium & High & P2P/All-to-All & ToR-Agg & Medium \\
 & EP & Medium & High & All-to-All & ToR-Agg & Medium \\ \hline
\multicolumn{1}{l|}{\multirow{2}{*}{\textbf{\begin{tabular}[c]{@{}l@{}}Inter-\\ Stage\end{tabular}}}} & WS & High & Low & T2G & ToR-Agg-Core & High \\
\multicolumn{1}{l|}{} & RS & Low & Medium & G2T & ToR-Agg-Core & High \\ \hline
\multirow{3}{*}{\textbf{\begin{tabular}[c]{@{}c@{}}Gen.\\ Stage\end{tabular}}} & TP & Medium & High & AllReduce & HBD & Low \\
 & EP & Medium & High & All-to-All & ToR-Agg & Low \\
 & AFD & Low & High & M2N (Bipartite) & ToR-Agg & Low \\ \hline
\end{tabular}%
}
\caption{Communication profiles of the disaggregated RL workflow on a Clos-style fabric. For each operation, we summarize its dominant pattern (volume, frequency, primitives), which determines its placement scope (Possible Domain) and potential benefit from OCS reconfiguration over the EPS substrate (Reconfigurability). WS denotes weight synchronization and RS denotes response streaming; T2G and G2T denote \train to \gen and \gen to \train transfers.}
\label{tab:traffic_profiles}
\end{table}

\section{Dynamic Network Orchestration}
\label{sec:network}

Disaggregated RL creates stage-dependent time-varying communication that requires an adaptive network topology as established in \Cref{sec:communication_characterization}. 
Our adaptive compute orchestration (\Cref{sec:compute_orchestration}) further incurs dynamic traffic.  
\sys complements compute scheduling with an adaptive fabric (\fabric) that provides (i) stage-aware topology reconfiguration and (ii) slack-aware orchestration that schedules circuit reconfiguration within each stage's tolerance.

\subsection{Overview}
\label{subsec:fabric}

\begin{figure}[t]
  \centering
\includegraphics[width=0.445\textwidth]{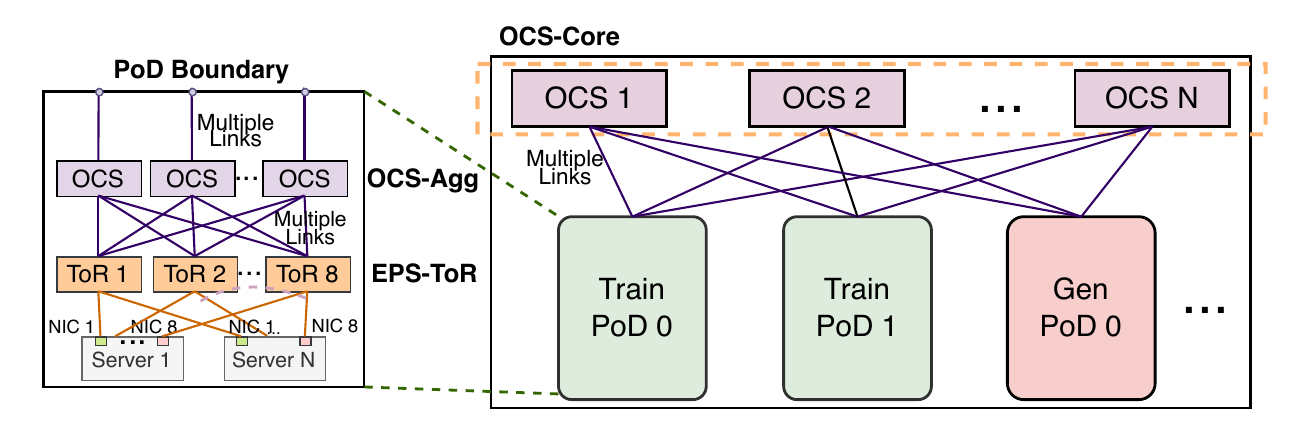}
  \vspace{-0.08in}
\caption{\fabric overview. Packet switching is performed at ToRs for host-facing connectivity and intra-ToR traffic. Above the ToR layer, aggregation and core connectivity are provided by reconfigurable OCS circuits.}
\label{fig:network_overview} 
\end{figure}

We adopt a hybrid EPS--OCS fabric called \fabric as depicted in Figure~\ref{fig:network_overview}. For each stable communication period
(within an RL stage), \sys selects a stage-aware topology template (\Cref{subsec:stage_aware_reconfig}) and proactively materializes the required OCS
circuits in the aggregation/core layers, while the underlying host-facing packet-switched substrate remains unchanged. Reconfiguration timing is
governed by \emph{safe windows} that can fully overlap the update cost; we detail this proactive orchestration in \Cref{subsec:orchestration}.

\fabric enforces a clear division of labor:
\begin{itemize}[leftmargin=*]
   \item \textbf{Access layer: Static EPS-ToR (rail-aligned).}
     As shown in Figure~\ref{fig:network_overview}, servers connect exclusively to static EPS ToRs. We employ a rail-aligned strategy~\cite{rail-optimized} here with compatibility for standard fat-trees. This electrical termination preserves stable host-facing links with low latency.

   \item \textbf{Aggregation \& core layers: Dynamic OCS.}
    Above the ToR, \fabric uses reconfigurable OCS:
    \begin{itemize}
        \item \textbf{OCS-Agg (intra-PoD):} Aggregation-layer OCS devices interconnect ToRs within a PoD, providing on-demand
        bandwidth for intra-PoD collectives (e.g., EP all-to-all) without changing any host-facing links.
        \item \textbf{OCS-Core (inter-PoD):} Core-layer OCS devices interconnect PoDs, materializing inter-PoD paths only when needed.
    \end{itemize}
\end{itemize}

\noindent Operationally, \sys steers traffic by scope. Traffic within a ToR uses the always-on EPS fabric. Traffic that spans ToRs (within a PoD) or PoDs is routed over the optical circuits accordingly. By reshaping only the upper-layer topology, \sys provides on-demand bandwidth without provisioning a worst-case static electrical core.

\begin{figure*}[t]
  \centering
  \includegraphics[width=0.865\textwidth]{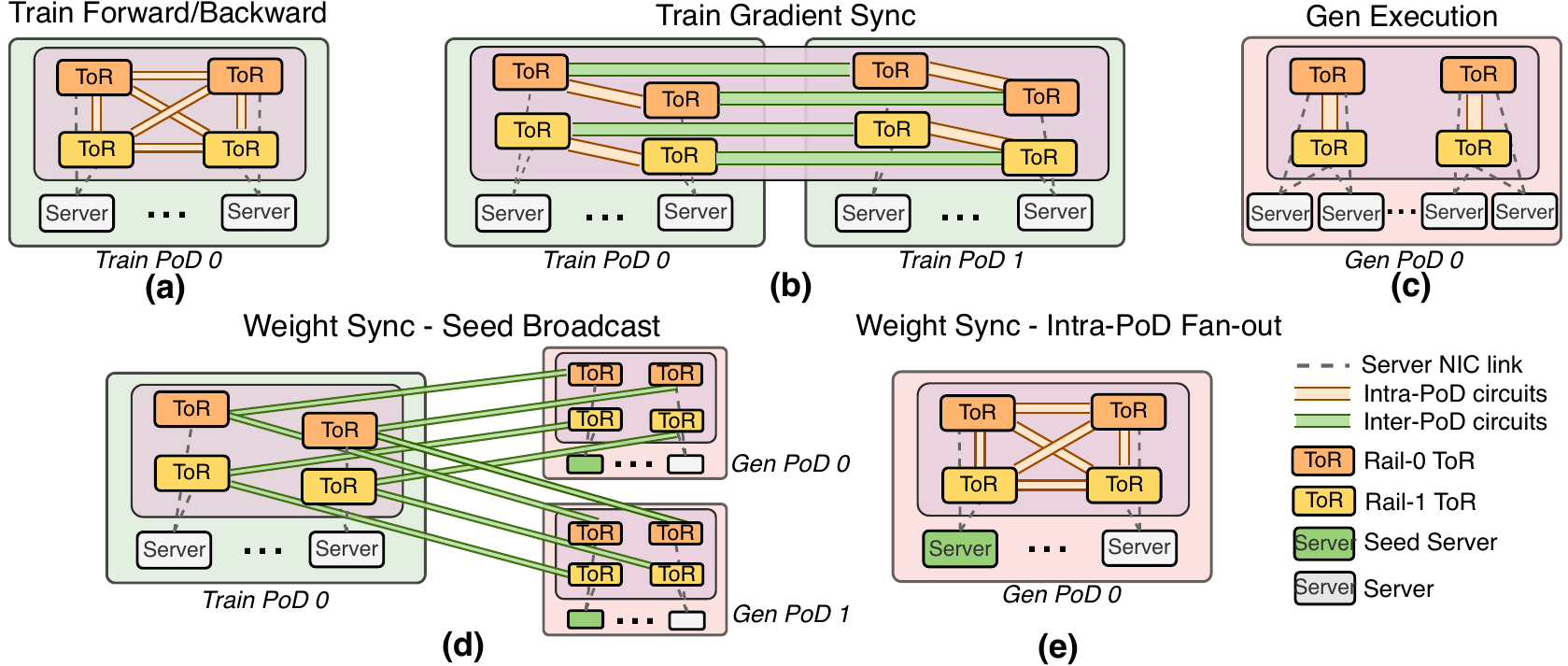}
  \vspace{-0.12in}
\caption{Dynamic topology materialization with \fabric across execution periods in different RL stages.
Each PoD contains multiple servers; for clarity, each server is dual-homed with two NICs connected to rail-0 and rail-1 ToRs.
Colored links denote OCS optical circuits (intra-PoD and inter-PoD, as indicated in the legend).
During \train forward/backward, the fabric may reconfigure across model layers as the dominant communication pattern changes; we show a representative snapshot.}
\label{fig:network_case} 
\end{figure*}

\subsection{Stage-Aware Topology Reconfiguration}
\label{subsec:stage_aware_reconfig}

\sys operates \fabric via \emph{dynamic topology materialization}: it re-configures optical circuits on-demand to instantiate a topology tailored to each stable communication period within an RL stage. We now describe (i) reconfiguration granularity across stages and (ii) the topology templates used
for \train, \gen, and weight synchronization.

\noindent\textbf{Reconfiguration granularity across RL stages.}
 \sys adapts its reconfiguration frequency based on available slack across RL stages.
 
 During \train, non-communication segments are often longer (e.g., due to large-batch execution), providing larger overlap opportunities. As a result,
 \sys can occasionally reconfigure within an iteration at selected stage-internal collective-free boundaries where the dominant primitive or the
 participant set changes (e.g., attention-dominated vs.\ FFN/MoE), and at the transition from computation to gradient synchronization.
 During \gen, overlap is limited, so \sys uses coarse-grained reconfiguration. It typically keeps one circuit plan for each interval between deployment boundaries (parallelism/layout changes), selecting a template from the interval's aggregated demand, provisioning OCS-Agg accordingly, and holding the plan until the next boundary.
 Finally, Weight Sync often provides ample non-critical slack, enabling on-demand reconfiguration right before the broadcast without extending
 the critical path.

\noindent\textbf{Stage-specific topology templates.} Figure~\ref{fig:network_case} illustrates how \sys materializes distinct topologies to match the spatial heterogeneity
characterized in \Cref{sec:characterization}, using the requirements summarized in Table~\ref{tab:traffic_profiles}.

\noindent\textbf{1) High-bisection fabric for \train.}
Most \train traffic during forward/backward is intra-PoD.
Accordingly, \fabric uses OCS-Agg to increase ToR-to-ToR bisection within a PoD (Figure~\ref{fig:network_case}(a)),
sustaining layer-wise collectives such as All-to-All across servers.
For DP AllReduce (gradient synchronization), which could span the whole \train clusters domain (Table~\ref{tab:traffic_profiles}), \sys additionally configures
OCS-Core circuits to form a high-bisection inter-PoD mesh (Figure~\ref{fig:network_case}(b)).
Although gradient exchange can be heavy for large models, it typically has limited overlap with other \train collectives,
allowing these circuits to be time-shared efficiently.

\noindent\textbf{2) Isolated intra-PoD fabrics for \gen.}
While \gen can be dominated by intra-server computation/communication (e.g., TP within a node), the \emph{network-critical}
portion is primarily localized cross-server traffic within each PoD (e.g., EP All-to-All~\cite{guo2025deepseekr1} and
M2N under AFD~\cite{zhu2025megascale,wang2025step}).
As summarized in Table~\ref{tab:traffic_profiles}, this traffic largely resides in the ToR--Agg domain.
Accordingly, \fabric leverages OCS-Agg to carve out independent intra-PoD topologies
(Figure~\ref{fig:network_case}(c)), isolating \gen PoDs from each other and avoiding over-provisioning the core for
traffic that rarely leaves a PoD.
In our example, we reserve only a small slice of core bandwidth to connect \gen PoDs to the core, which is
sufficient to stream generated responses back to \train PoDs.

\noindent\textbf{3) Tree-style broadcast for weight synchronization.}
To accelerate periodic weight synchronization---an infrequent but high-volume transfer spanning many servers
(Table~\ref{tab:traffic_profiles})---\fabric materializes a temporary, circuit-scheduled dissemination \emph{forest}
using OCS-Core and OCS-Agg (Figure~\ref{fig:network_case}(d--e)), where each weight shard is disseminated via its own
tree.
The design is \emph{two-level}. In Phase~A, OCS-Core provisions inter-PoD point-to-point circuits from shard holders in \train PoDs to a small set of seed receivers in each destination \gen PoD, which serve as ingress points and fanout roots within the PoD. 
Each inter-PoD transfer also uses intra-PoD connectivity at both ends to reach boundary-facing uplinks
(via OCS-Agg and rack-local EPS).
Phase~A executes in multiple waves under a greedy, budget-aware scheduler that respects per-PoD boundary concurrency
budgets (i.e., a cap on simultaneously active boundary-facing circuits incident to each PoD), preventing boundary-port
hotspots.
In Phase~B, within each destination PoD, \fabric uses OCS-Agg (or rack-local EPS) to provision intra-PoD fanout from
seeds to all remaining \gen ranks, avoiding redundant replication across the PoD boundary and alleviating root-NIC/ToR
bottlenecks.
Appendix~\ref{app:weightsync-tree} details the overlay construction and scheduling.

\subsection{Orchestration for Proactive Reconfiguration}
\label{subsec:orchestration}

The runtime triggers reconfiguration only when the next \emph{safe window}---a collective-free slack interval identified at runtime---is expected to cover the OCS update overhead at the chosen granularity; otherwise, \sys retains the current plan for that period. Building on the stage-aware templates (\Cref{subsec:stage_aware_reconfig}), \sys programs OCS devices via a dedicated control proxy running on an out-of-band control server/network. The proxy applies updates only within safe windows, using explicit runtime signals from lightweight instrumentation of collective boundaries. For each upcoming communication period, the runtime supplies an \emph{intent} descriptor (stage type, dominant primitive mix, and group membership) and the expected transfer volume; the proxy materializes and commits the corresponding circuit plan in the next safe window (Algorithm~\ref{alg:topology_materialization}).

The proxy operates in two phases:

\noindent \textbf{{Phase 1:} Caching templates and allocation procedures.}
In the first iteration of an RL job, the proxy identifies recurring communication periods at the per-stage
reconfiguration granularity. For each period, it stores the intent descriptor and caches (i) the selected topology
template and (ii) a demand-driven allocation procedure (Appendix~\ref{app:portalloc}). At runtime, the cached procedure
instantiates a concrete circuit plan from the current demand and fabric state (e.g., free ports and masked failures).

\noindent \textbf{{Phase 2:} Proactive reconfiguration with runtime adaptation.}
In subsequent iterations, given the cached template/procedure and runtime-announced participants/volumes, the proxy
materializes a concrete circuit plan for the upcoming period. Using a ToR-centric view, it constructs demand at two
scopes: (i) ToR-level demands within each PoD for
ToR--ToR (intra-PoD) and ToR--uplink (inter-PoD lift/drop) circuits and (ii) a PoD-to-PoD summary to size OCS--Core inter-PoD capacity. The proxy then runs a greedy allocator
(Appendix~\ref{app:portalloc}) that iteratively adds one full-duplex circuit to the eligible pair with the largest
\emph{demand-to-capacity ratio} (i.e., demand divided by currently allocated circuit capacity), subject to port
availability, until no demanded pair has free ports at both ends. 
Before committing, it performs a final end-to-end feasibility check (e.g., Core--Agg coupling and any state changes since materialization); if the check fails, it keeps the previous plan and retries in a later safe window.

\begin{algorithm}[t]
\caption{Topology Materialization}
\footnotesize
\label{alg:topology_materialization}
\begin{algorithmic}[1]
\State \textbf{Input:} demand $D$ (for the upcoming communication period),
intent $\mathit{I}$ (stage type / primitive mix / group membership),
fabric state $S$ (free ports, masked failures, link bandwidth $B_{\text{link}}$, previous plan $C_{\text{prev}}$)
\State \textbf{Output:} active circuit plan $C_{\text{act}}$

\State $tpl \gets \text{GetCachedTemplate}(\mathit{I})$ \Comment{Populate on first iteration if missing}
\State $G \gets \text{AggregateDemand}(D,\ \mathit{I})$ \Comment{PoD-level (core) and/or ToR-level (agg)}
\State $C \gets \text{AllocateCircuits}(tpl,\ G,\ S)$ \Comment{See Appendix~\ref{app:portalloc}}
\If{$\neg \text{Validate}(C,\ S)$ }
  \State \Return $C_{\text{prev}}$ \Comment{Keep the last plan}
\EndIf
\State \textsc{CommitInNextSafeWindow}$(C)$
\State \Return $C$
\end{algorithmic}
\end{algorithm}

\subsection{Practical Considerations}

\noindent\textbf{Non-blocking reconfiguration.} 
During topology reconfiguration, \fabric maintains switch-to-switch connectivity while dynamically adjusting the number of active links to modulate bandwidth.
Even if the topology materialization latency exceeds the required reconfiguration window, communication operations can proceed without stalling for the reconfiguration completion.
Moreover, with the assistance of ECMP and re-routing, the available bandwidth for communication changes smoothly, at the cost of only a limited number of packet drops and reorders. 
Consequently, jobs can proceed with reduced, though not optimal, bandwidth until all switched links are fully activated.

\noindent\textbf{Failure handling.}
\fabric is robust to device and link failures. Optical switching does not introduce new qualitative data-plane failure modes; it mainly changes which circuits are provisioned and the available bandwidth. 
\fabric terminates OCS at the ToR/aggregation layers, keeping host-facing EPS--ToR links always on and avoiding endpoint NIC/transceiver re-activation overheads seen with NIC-terminated OCS~\cite{liao25mixnet}.
It handles failures at three layers:
\emph{(i) Server failures:} upon a GPU server failure, the job replaces failed ranks, and \fabric
re-materializes optical connectivity to match the updated placement.
\emph{(ii) OCS failures:} if an OCS port/circuit is unavailable, the proxy masks the failed resources and
reallocates circuits under the remaining port budget, typically reducing bandwidth but preserving connectivity.
\emph{(iii) EPS/NIC failures:} if an EPS link attached to a NIC fails, traffic can be relayed over the intra-server
scale-up fabric (e.g., NVLink) to a healthy NIC and exit via its EPS links. We evaluate their performance impact in Appendix~\ref{app:network_failure}.

\section{Testbed Evaluation}

\noindent\textbf{Setup.} We conducted experiments on an H800 testbed with servers interconnected by an 8$\times$200Gbps RDMA network with a rail-optimized topology to validate the effectiveness of \sys's compute scheduler. Adapted from veRL~\cite{sheng2025verl}, we use Megatron-LM~\cite{shoeybi2019megatron} as the training framework and a vLLM-based~\cite{kwon2023vllm} backend for generation. 

\noindent\textbf{Baselines.} We compare \sys against the following baselines of one-step asynchronous RL:
\begin{itemize}[leftmargin=*]
    \item \noindent\textbf{veRL-TO.} This baseline prioritizes DP to maximize the number of concurrent generation instances, tuning the DP-TP combination through a parameter sweep (TP = 1, 2, 4, 8) to achieve the best overall performance.
    \item \noindent\textbf{veRL-LO.} This baseline maximizes TP=8 for each generation instance to minimize latency.
    \item \noindent\textbf{Partial-rollout (PR).} An extension of veRL-TO, this approach enables partial rollouts, allowing a single response to be processed across two consecutive model versions. In this mode, a response can be truncated at one step and resumed in a subsequent step, though this may introduce greater staleness.
    
\end{itemize}

\noindent\textbf{Workloads.}  We evaluate the Qwen-2.5 14B and 32B models~\cite{qwen2025qwen25technicalreport} on the openr1-math-220k~\cite{OpenR1-Math-220k} and deepmath-103k~\cite{he2025deepmath} datasets, respectively, using the GRPO algorithm~\cite{shao2024deepseekmathgrpo}. Model-related details are provided in Table~\ref{tab:qwen25_configs} of Appendix~\ref{app:model_config}.

\subsection{End-to-End Performance}

\begin{figure}[t]
  \centering
\includegraphics[width=0.485\textwidth]{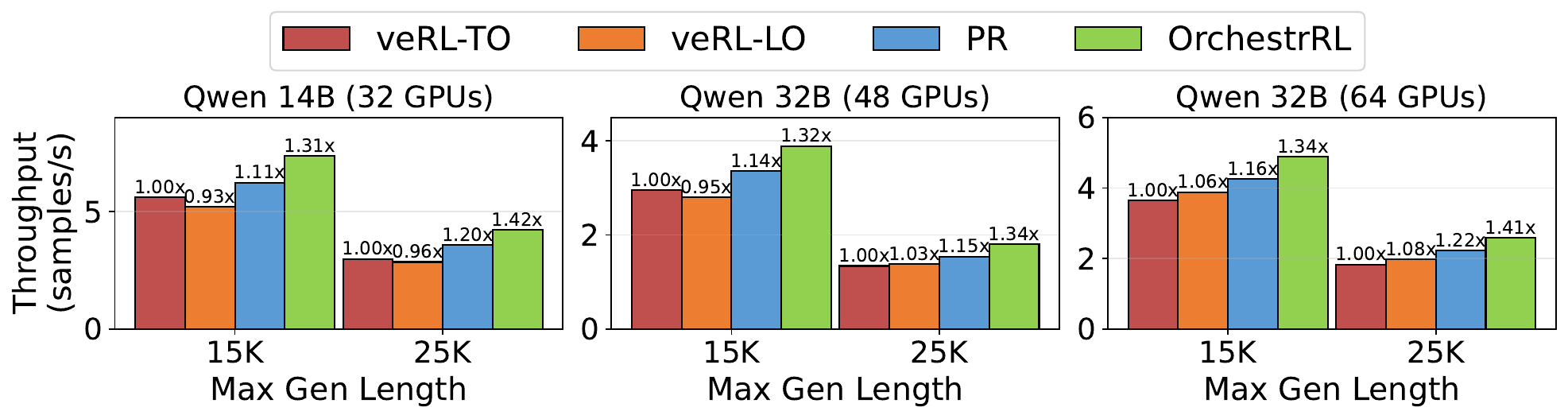}
  \vspace{-0.25in}
\caption{End-to-end throughput comparison across different schemes.}
\label{fig:end_to_end_throughput} 
\end{figure}

Figure~\ref{fig:end_to_end_throughput} reports the end-to-end throughput, defined as the average number of samples processed per second. \sys consistently outperforms all baselines. For Qwen-14B on 32 GPUs and Qwen-32B on 48 GPUs, we provision the training and generation clusters with the same number of GPUs. On 32 GPUs, \sys improves throughput over veRL-TO by 1.31$\times$ at a maximum generation length of 15K tokens and 1.42$\times$ at 25K tokens. On 48 GPUs with Qwen-32B, \sys achieves similar gains of 1.32$\times$ at 15K tokens and 1.34$\times$ at 25K tokens. We observe the same trend for Qwen-32B on 64 GPUs (with 40 GPUs allocated to generation). Overall, these results highlight the benefit of \sys's adaptive compute scheduling, consistent with the case study in \Cref{exp:case_study}.

\subsection{Ablations and Analysis} 

\subsubsection{\bf \em  Ablation study}

\begin{figure}[t]
  \centering
  \includegraphics[width=0.455\textwidth]{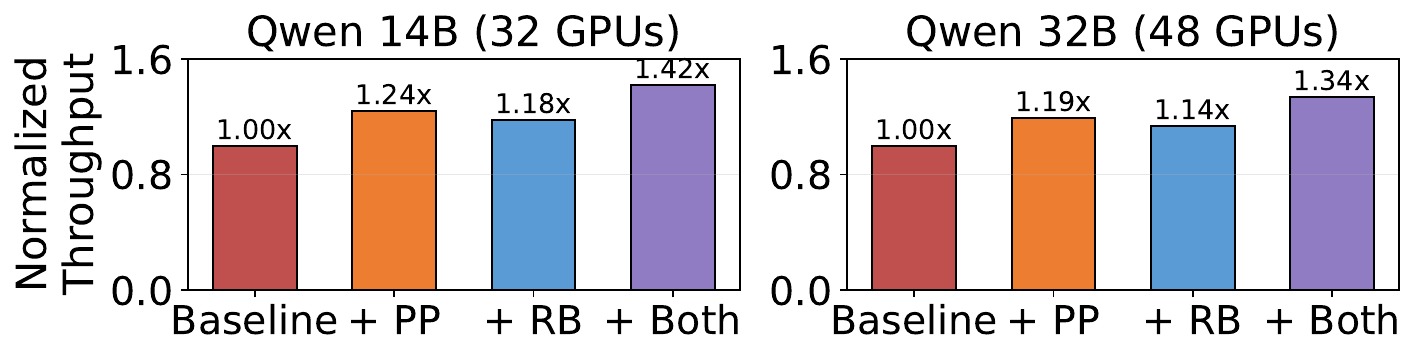}
  \vspace{-0.15in}
  \caption{Improvement breakdown of \sys.}
  \label{fig:breakdown}
\end{figure}

Proactive planning (+PP) adapts parallelism within a generation step to the evolving workload, improving throughput to 1.24$\times$ on Qwen-14B and 1.19$\times$ on Qwen-32B over veRL-TO (Figure~\ref{fig:breakdown}). Reactive balancing (+RB) is also beneficial (1.18$\times$/1.14$\times$), using lightweight on-the-fly request migrations to mitigate output-length variability and KV-cache imbalance. Combining the two delivers the largest gains, reaching 1.42$\times$ and 1.34$\times$.

\subsubsection{\bf \em Case study}\label{exp:case_study}

\begin{figure}[t]
  \centering
  \includegraphics[width=0.315\textwidth]{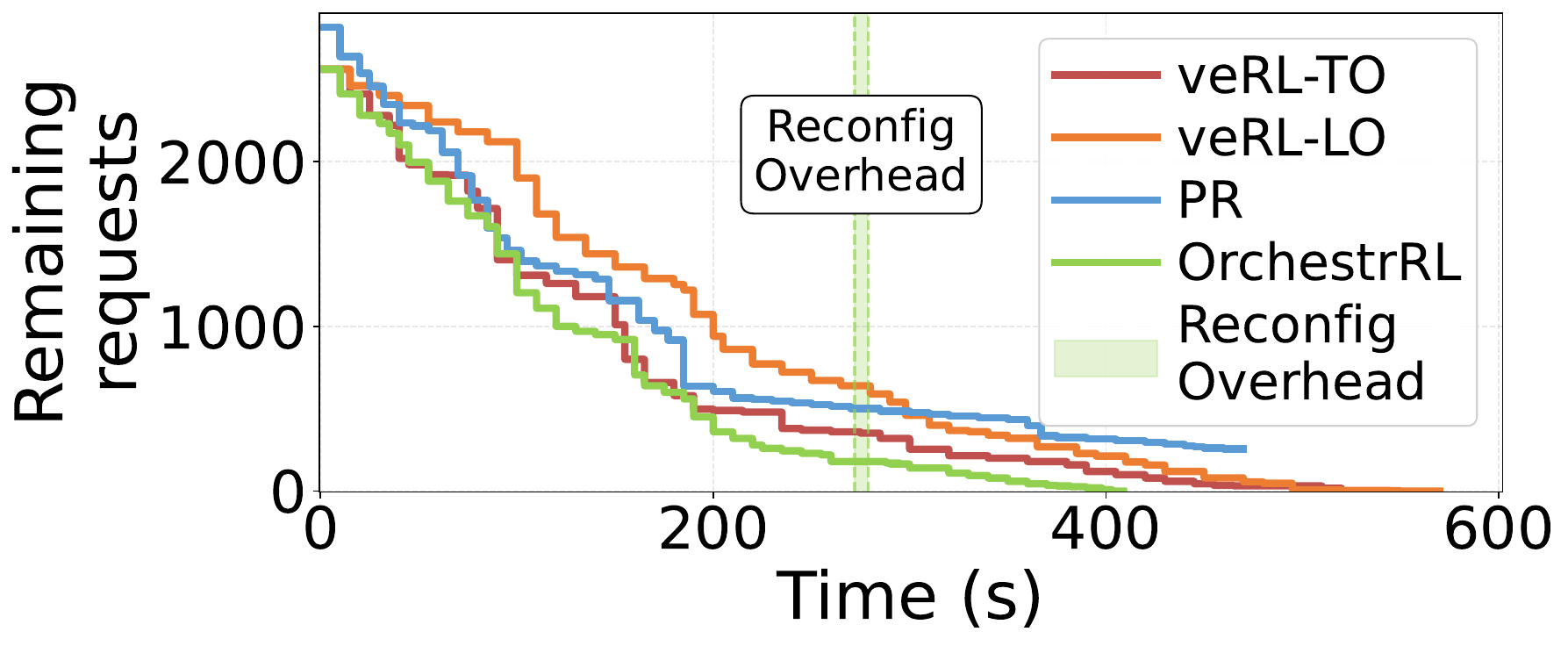}
  \vspace{-0.15in}
  \caption{Remaining requests in a generation step across different schemes.}
  \label{fig:remaining_requests_with_config}
  \vspace{-3mm}
\end{figure}

We present remaining-request curves for Qwen-14B in Figure~\ref{fig:remaining_requests_with_config}.
veRL-LO starts with low concurrency (a few TP=8 instances), limiting early parallelism and slowing progress.
veRL-TO uses eight TP=2 instances, clearing the initial batch quickly but becoming less efficient as concurrency drops; it still finishes faster than veRL-LO.
PR starts with more remaining requests due to truncation from the previous step and ends with truncation under its rollout policy.
\sys adapts to the workload by balancing requests and reconfiguring parallelism, switching from eight TP=2 to two TP=8 instances to shorten the tail.
Although reconfiguration adds \(\sim\)7.5s overhead (mainly 1.5s for switching and 5.4s for KV-cache recomputation) at around 272s, \sys achieves the shortest completion time. Optimization solving time is negligible in this setting; detailed results are shown in Figure~\ref{fig:solve_time} (Appendix~\ref{app:optim_solve_time}).

\section{Large-Scale Simulation}
To evaluate \fabric and the large-scale behavior of \sys, we build {RLSim}, a high-fidelity simulator for disaggregated RL at cluster scale. {RLSim} integrates (i) a training simulator, (ii) an inference simulator based on Frontier~\cite{feng2025frontier}, and (iii) a packet-level network simulator extended from htsim~\cite{htsim}. These components model the end-to-end RL pipeline, support multiple network topologies, and capture diverse collective communication patterns.

We use {RLSim} to (1) validate fidelity by matching per-stage training, inference, and network behaviors against measurements, (2) evaluate \fabric's performance-cost trade-offs against alternative fabrics, and (3) co-simulate \sys by integrating its compute scheduler with \fabric, reporting end-to-end throughput and throughput per dollar. We primarily evaluate a dense Qwen-2.5-72B model~\cite{qwen2025qwen25technicalreport} and the Qwen2-57B-A14B MoE model~\cite{qwen2-57B-A14B}. We use input- and output-length traces collected on our testbed with deepmath-103k~\cite{he2025deepmath} and scale them to match the simulated GPU count. Detailed model configurations are provided in Appendix~\ref{app:model_config}.

\subsection{Simulator Fidelity}

\begin{figure}[t]
  \centering
  \includegraphics[width=0.465\textwidth]{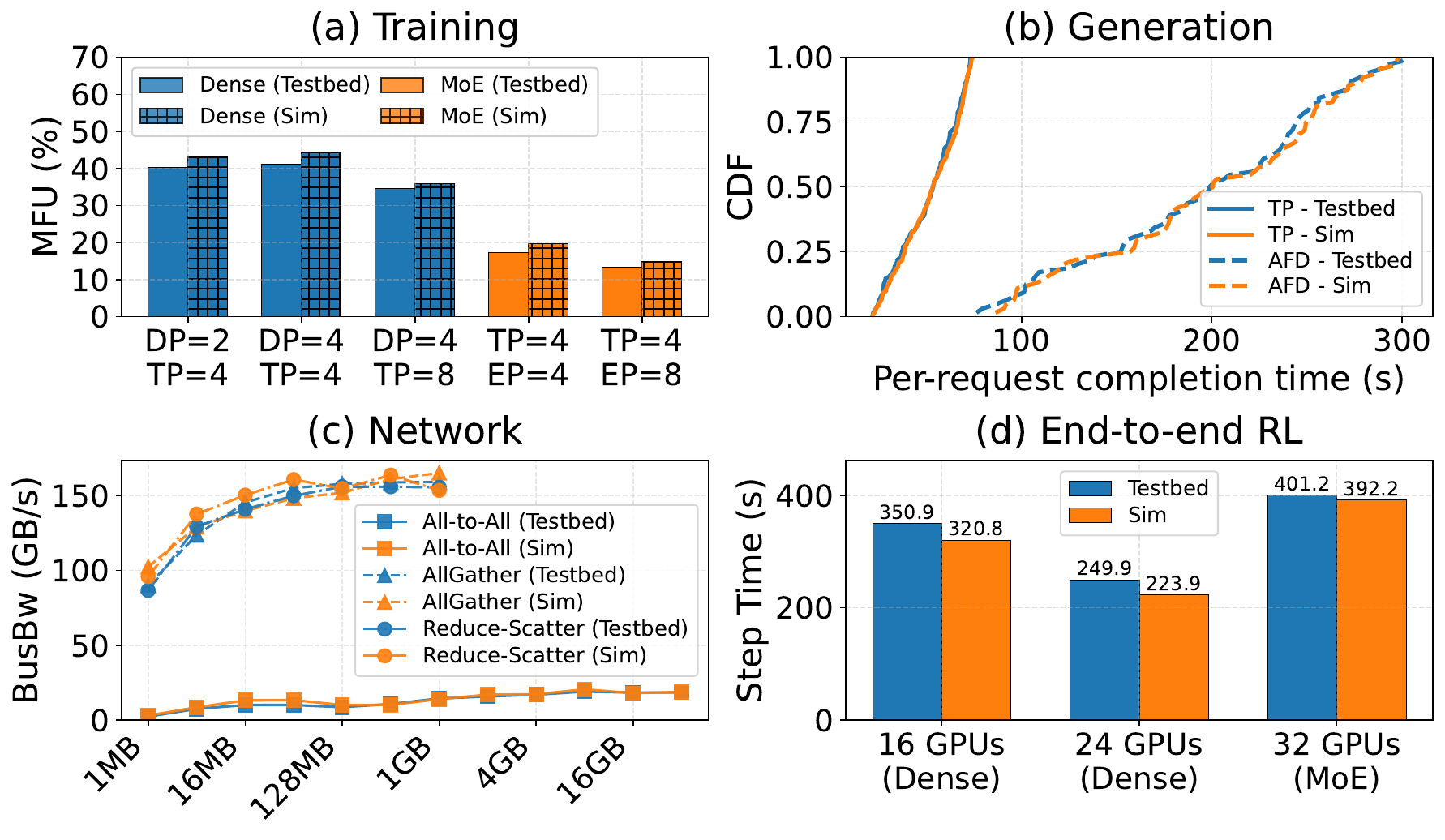}
  \vspace{-0.20in}
\caption{RLSim fidelity evaluation. In (c), the all-gather data size refers to the input tensor size, while the reduce-scatter data size refers to the output tensor size.}
\label{fig:simulator_fidelity} 
\vspace{-5mm}
\end{figure}

Here, we evaluate the fidelity of RLSim by validating its key components as well as the end-to-end workflow.

\noindent\textbf{Training.}
We evaluate fidelity by comparing model FLOP utilization (MFU) under a fixed workload (global batch size 512, sequence length 5120). We sweep DP/TP/EP for dense and MoE training on 8--32 GPUs (Figure~\ref{fig:simulator_fidelity}(a)). RLSim closely matches the testbed MFU levels and trends, with at most 8.86\% error.

\noindent\textbf{Generation.} We assess RLSim's inference fidelity with batch size 256, prompt length 512, and generation lengths from 1024 to 4096. Figure~\ref{fig:simulator_fidelity}(b) plots the completion time CDFs, showing that RLSim closely tracks the testbed result: the mean error is 3.3\% for TP and 5.95\% for AFD.\footnote{{Due to the lack of publicly available AFD implementation, we are only able to use an internal codebase with an internal MoE model for calibration.}}

\noindent\textbf{Network.}
We validate the packet-level network simulator by calibrating it to our rail-optimized fabric and replaying NCCL-style collective microbenchmarks across message sizes. We compare bus bandwidth on a 5-server (40-GPU) testbed against RLSim (Figure~\ref{fig:simulator_fidelity}(c)). RLSim matches the testbed within 5.98\% average error.

\noindent\textbf{End-to-end RL.} Finally, we validate end-to-end fidelity by measuring per-step wall-clock time of a full RL iteration (Figure~\ref{fig:simulator_fidelity}(d)). Across different configurations for dense and MoE models, RLSim achieves errors of 2.2\% to 10.4\%.

\subsection{Network-Centric Results }

\noindent\textbf{Setup.} 
Unless otherwise specified, we set the per-link propagation latency to \SI{1}{\micro\second} and the per-hop switch processing latency to \SI{0.8}{\micro\second}, following~\cite{liao25mixnet}. We use an MTU of 9216\,B and simulate packet-level queuing and congestion with identical NIC, link, and switch parameters across all evaluated topologies to ensure a fair comparison. Each server hosts eight NVIDIA H800 GPUs and eight NICs (one per GPU); each NIC provides bandwidth $B$ (default 200\,Gbps). Intra-server GPUs are connected via NVLink, providing an effective bidirectional bandwidth of 400 GB/s per GPU. For OCS-based designs, we model topology-dependent circuit connectivity and routing, assuming a 25 ms reconfiguration overhead unless specified otherwise. With a given budget, we allocate the number of GPUs to \train and \gen following~\cite{zhong2025streamrl}.

\noindent\textbf{Baselines.} We compare the performance of \fabric with the following interconnects:
\begin{itemize}[leftmargin=*]
    \item \noindent\textbf{Fat-tree (FT)}~\cite{al2008fattree}. A non-blocking Fat-tree network.
    \item \noindent\textbf{OverSub. Fat-tree (FT-OS).} A Fat-tree interconnect with the 3:1 over-subscription ratio.
    \item \noindent\textbf{Rail-optimized (RO)}. A non-blocking Fat-tree variant that connects same-rank GPUs to the same ToR switch, reducing intra-rail latency~\cite{rail-optimized}.
    \item \noindent\textbf{TopoOpt}~\cite{wang2023topoopt}. An optical interconnect that all NICs are optimistically connected directly via optical patch panels, which share a similar topology with \cite{khani2021sipml}.
    \item \noindent\textbf{MixNet}~\cite{liao25mixnet}. A state-of-the-art hybrid EPS-OCS fabric: in each 8-GPU server, 6 GPUs connect to a local OCS, while 2 provide scale-out connectivity via a non-blocking Fat-tree.

\end{itemize}

\begin{figure}[t]
  \centering
  \includegraphics[width=0.45\textwidth]{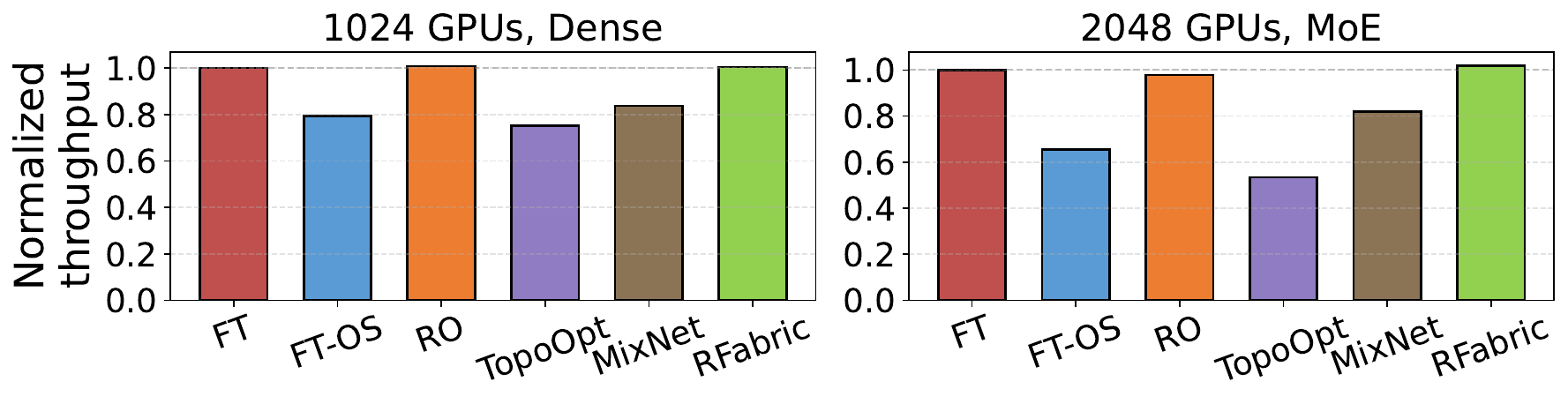}
  \vspace{-0.18in}
\caption{End-to-end performance comparison across different scales.}
\label{fig:network_performance} 
\end{figure}

\subsubsection{\bf \em End-to-end performance.}

Figure~\ref{fig:network_performance} illustrates significant throughput losses for FT-OS and TopoOpt. FT-OS is constrained by oversubscription, while TopoOpt struggles to provide sufficient scale-out bandwidth for \train-time gradient synchronization and \train/\gen weight synchronization. Its performance further degrades for bandwidth-sensitive collectives (e.g., all-to-all) when host-level forwarding is required, achieving only 0.47--0.76$\times$ of FT, with worse results on MoE models. Similarly, MixNet underperforms due to limited scale-out connectivity for DP traffic and large-scale weight synchronization, reaching approximately 0.81$\times$ of FT. In contrast, \fabric remains close to FT and RO, leveraging its flexible hybrid topology and stage-aware reconfiguration across RL stages.

\subsubsection{\bf \em Stage-wise performance breakdown}

\begin{figure}[t]
  \centering 
  \includegraphics[width=0.485\textwidth]{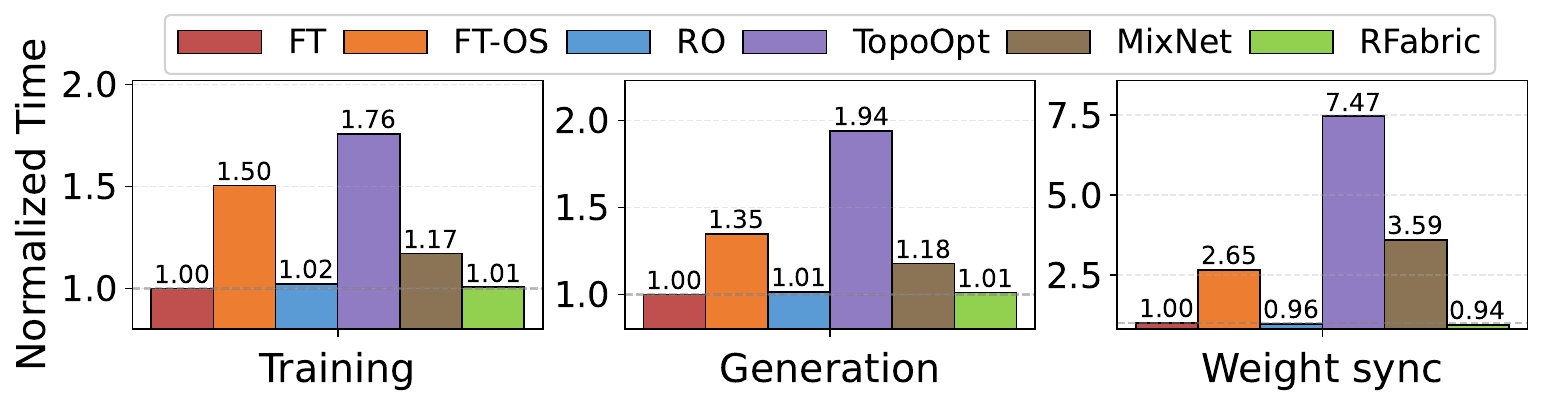}
  \vspace{-8mm}
\caption{Stage-wise performance under different network fabrics for Qwen-57B-MoE model.}
\label{fig:stage_wise_performance} 
\vspace{-3mm}
\end{figure}

Figure~\ref{fig:stage_wise_performance} reports the per-stage time of Qwen2.5-57B-MoE across \train, \gen (AFD mode), and \weightsync, normalized to FT. \fabric stays close to FT across all stages, matching RO and yielding a small improvement in \weightsync by leveraging optical circuits for bulk transfers. In contrast, FT-OS, TopoOpt, and MixNet are consistently slower to varying degrees.

\subsubsection{\bf \em Network cost}

\begin{figure}[t]
  \centering
  \includegraphics[width=0.48\textwidth]{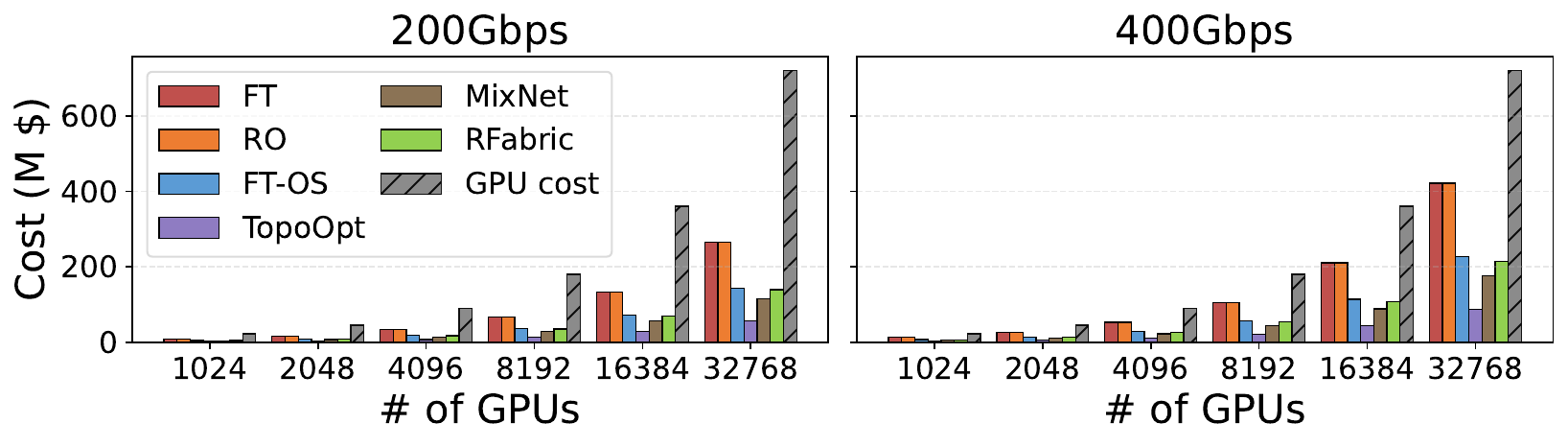}
  \vspace{-0.28in}
\caption{Network cost comparison at 200Gbps and 400Gbps. GPU cost is shown as reference. Detailed pricing is provided in Appendix~\ref{app:network_price}.}
\label{fig:network_cost} 
\vspace{-4mm}
\end{figure}

We analyze network costs for clusters with 200Gbps and 400Gbps links (Figure~\ref{fig:network_cost}). The FT is the high-cost baseline (1.00$\times$), and Rail-Optimized has similar cost. Oversubscribed FT reduces cost to 0.54$\times$, while TopoOpt further lowers it to 0.21$\times$ at the expense of reduced scale-out bandwidth for global collectives. In contrast, \fabric costs 0.51--0.52$\times$ of FT, comparable to MixNet and FT-OS; this relative ordering is consistent across link speeds. Importantly, although GPU cost remains higher than network cost in absolute terms, faster links increase the network-to-GPU cost ratio: for FT, network cost grows from $\sim$37\% of GPU cost at 200Gbps to $\sim$59\% at 400Gbps, making network cost control increasingly important at higher link rates.

\subsubsection{\bf \em Performance-cost pareto frontier}

\begin{figure}[t]
  \centering
\includegraphics[width=0.47\textwidth]{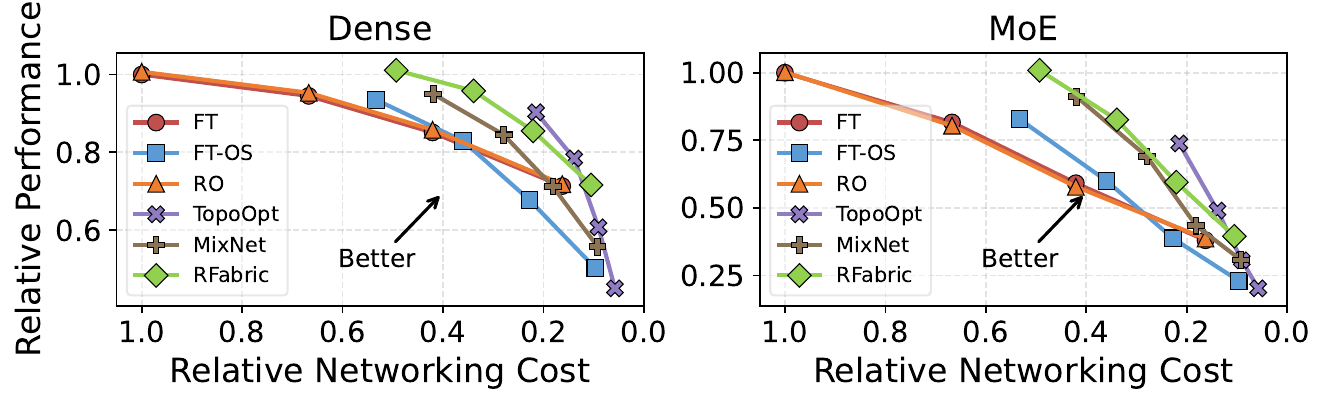}
  \vspace{-0.14in}
\caption{Performance-cost trade-off under different interconnect link speeds. For each curve, the four markers from left to right correspond to 800/400/200/100 Gbps.}
\label{fig:network_pareto_frontier} 
\vspace{-5mm}
\end{figure}

Figure~\ref{fig:network_pareto_frontier} summarizes the performance-cost trade-off under four link speeds. {\fabric} stays on or near the Pareto frontier, providing higher performance at a given cost (or lower cost at a given performance) than most alternatives. In particular, TopoOpt and FT-OS reduce cost but suffer substantial performance degradation, placing them well below the frontier. Compared with high-performance baselines, {\fabric} improves cost-efficiency by $1.53\times$--$2.06\times$ over FT and by $1.61\times$--$2.08\times$ over RO, and by $\sim$$1.13\times$ over MixNet. 

\subsubsection{\bf \em OCS-related sensitivity analysis.}
We evaluate sensitivity to (i) OCS reconfiguration delay $T_{ocs}$ and (ii) the circuit allocation policy (Appendix~\ref{app:sensitive}). As discussed in \Cref{sec:network}, $T_{ocs}$ impacts stages differently: \train can tolerate tens of milliseconds due to slack; \gen cannot reconfigure on the critical path and thus only reconfigures at coarse deployment boundaries; and \weightsync is largely insensitive over practical ranges given its ample slack. We also compare traffic-aware allocation with uniform splitting, finding the former superior as it focuses circuits on bottleneck pairs, while the latter wastes capacity on low-demand links.

\subsection{Co-Simulation}

\noindent\textbf{Setup.} We evaluate the compute-network co-design in \sys through a $2\times2$ factorial ablation study, considering two key factors: (i) compute scheduling (\sys scheduler vs.\ a fixed baseline) and (ii) network fabric (\fabric vs.\ a cost-comparable static Fat-tree network with $3{:}1$ oversubscription). This results in four configurations: \textit{Baseline}, \textit{Sched-only}, \textit{Fabric-only}, and \sys. We evaluate Qwen3-235B-A22B~\cite{Qwen3-235B-A22B} on 2048 GPUs.
\subsubsection{\bf \em  End-to-end results.}

\begin{figure}[t]
  \centering
  \begin{minipage}[t]{0.31\textwidth}
    \centering
    \includegraphics[width=\linewidth]{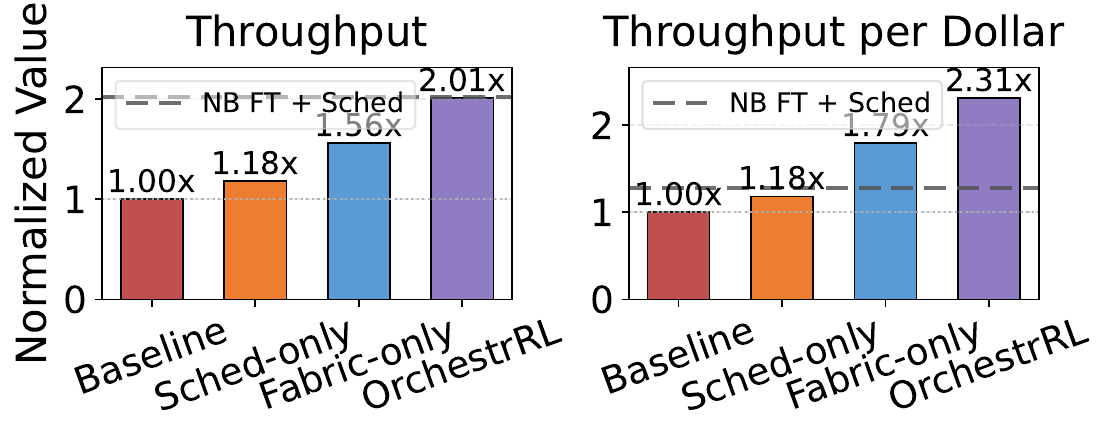}
    \vspace{-0.27in}
    \captionof{figure}{Co-design comparison. Dashed line is a reference setup of full Fat-tree and \sys's compute scheduler.}
    \label{fig:co_sim_results}
  \end{minipage}
  \hfill
  \begin{minipage}[t]{0.16\textwidth}
    \centering
    \includegraphics[width=\linewidth]{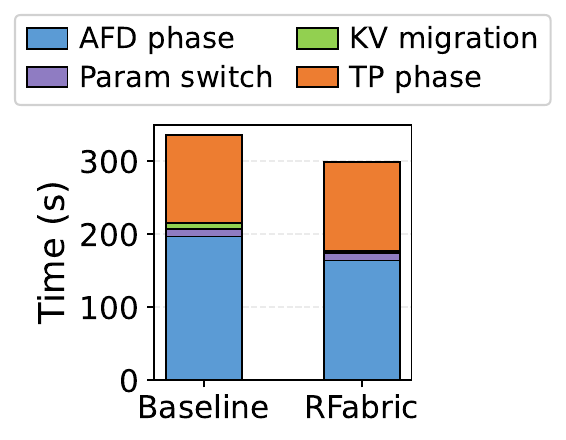}
    \vspace{-0.27in}
    \captionof{figure}{Network impact on compute scheduler in \gen.}
    \label{fig:network_impact}
  \end{minipage}
\vspace{-6mm}
\end{figure}

Figure~\ref{fig:co_sim_results} reports end-to-end throughput and throughput-per-dollar (TPD), both normalized to the baseline, where \(\mathrm{TPD} = \mathrm{Throughput} / \mathrm{Cost}\), with \emph{Cost} including total network cost.

\noindent\textbf{Throughput.}  
As shown in Figure~\ref{fig:co_sim_results}(a), \textit{Sched-only} achieves a modest \(1.18\times\) throughput gain, while \textit{Fabric-only} improves throughput to \(1.56\times\), underscoring the network bottleneck under the oversubscribed Fat-tree baseline. Combining both, \sys achieves the highest throughput (\(2.01\times\)), matching the non-blocking Fat-tree + scheduler reference.

\noindent\textbf{Cost efficiency.}  
Figure~\ref{fig:co_sim_results}(b) shows \textit{Sched-only} improves TPD by \(1.18\times\), while \textit{Fabric-only} achieves \(1.79\times\) by improving performance and reducing costs. \sys combines both benefits, delivering the best TPD (\(2.31\times\)) and outperforming the non-blocking Fat-tree + scheduler reference, which incurs significantly higher network cost.

\subsubsection{\bf \em Network's impact on compute reconfiguration.}
To quantify how the fabric interacts with \sys's compute scheduler during reconfiguration, we compare a representative \gen step on \fabric to a cost-comparable oversubscribed Fat-tree. The step transitions from 35 AFD instances (each using \(3\times8\) GPUs for attention and \(2\times8\) GPUs for MoE) to 175 TP\(=8\) instances, triggering model-parameter reloads and KV-cache migration across servers. With \fabric, the AFD phase time drops by 16.7\% and KV-cache migration by 58.3\%, while parameter switching and the TP phase remain essentially unchanged. Overall, \fabric reduces end-to-end \gen step time by 10.7\% versus the baseline.

\section{Related Work}

\noindent\textbf{RL training frameworks.}
Many frameworks accelerate RL training, spanning co-located designs and their optimizations~\cite{sheng2025verl,hu2024openrlhf,zhong2024rlhfuse} as well as newer asynchronous, disaggregated systems such as StreamRL and Laminar~\cite{hu2024openrlhf,sheng2025verl,fu2025areal,zhong2025streamrl,sheng2025laminar}. \sys targets one-step asynchrony to balance stability and throughput, focusing on the generation bottleneck and a co-designed network fabric.

\noindent\textbf{Generation optimization in RL.}
Generation is a key bottleneck in RL due to the long-tailed distribution of output lengths~\cite{gao2025rollpacker,zhong2025streamrl,wu2025llamarl,sheng2025laminar,longzhao2025phase}. Prior methods address this with speculative decoding~\cite{he2025history,qin2025seer} and partial rollouts~\cite{fu2025areal,gao2025rollpacker}, mostly in synchronous settings, though some~\cite{gao2025rollpacker,fu2025areal} incur additional staleness. \sys expands the design space by dynamically adapting parallelism and performing lightweight, on-the-fly load balancing, mitigating stragglers caused by variable output lengths and uneven KV-cache usage---without introducing extra staleness. Agentic RL with external tools needs to optimize per-request workflows~\cite{tan2025towards,li2025continuum}; \sys is complementary and can integrate into such pipelines, primarily targeting RL reasoning settings

\noindent\textbf{OCS-related network architectures.}
Prior hybrid packet/circuit DCNs primarily optimize for general traffic patterns, failing to capture the phase-varying collective communication in modern LLM/RL workloads. This often results in connectivity mismatches or reconfiguration misaligned with stage-level communication periods~\cite{bcube,jellyfish,VL2,portland,helix,hedera,poutievski2022jupiter,Urata2022Apollo}. Recent approaches~\cite{ding2025photonic,khani2021sipml,wang2023topoopt}, such as SiP-ML~\cite{khani2021sipml} and TopoOpt~\cite{wang2023topoopt}, co-optimize topology and parallelization but rely on centralized, one-shot OCS reconfiguration, limiting their ability to support large-scale heterogeneous parallelism (e.g., MoE). While MixNet~\cite{liao25mixnet} introduces regional reconfigurability for MoE training, it confines reconfiguration to local domains with GPU-terminated circuits, restricting cluster-wide bandwidth and flexibility for latency-sensitive tasks. In contrast, \fabric is the first OCS-based fabric designed explicitly for disaggregated RL. It enables phase-aware, multi-granularity reconfiguration across training, generation, and weight synchronization using higher-tier OCS connectivity, addressing the limitations of prior systems.

\section{Conclusion}

This paper presents \sys, an orchestration framework for disaggregated RL that mitigates inefficiencies caused by workload dynamics. \sys includes a compute scheduler that switches parallelism and balances requests during generation to reduce makespan. It also integrates \fabric, a reconfigurable EPS-OCS fabric that adapts topology to demand, avoiding network overprovisioning while sustaining performance. Our evaluation shows substantial gains in throughput and cost efficiency.

\clearpage
{
\bibliographystyle{plain}
\bibliography{main}
}

\clearpage
\twocolumn
\newpage

\appendix

\section{Cost of KV-Cache Migration vs. Recomputation}
\label{app:kv-migration-cost}

\begin{figure}[t]
    \centering
\includegraphics[width=0.9\linewidth]{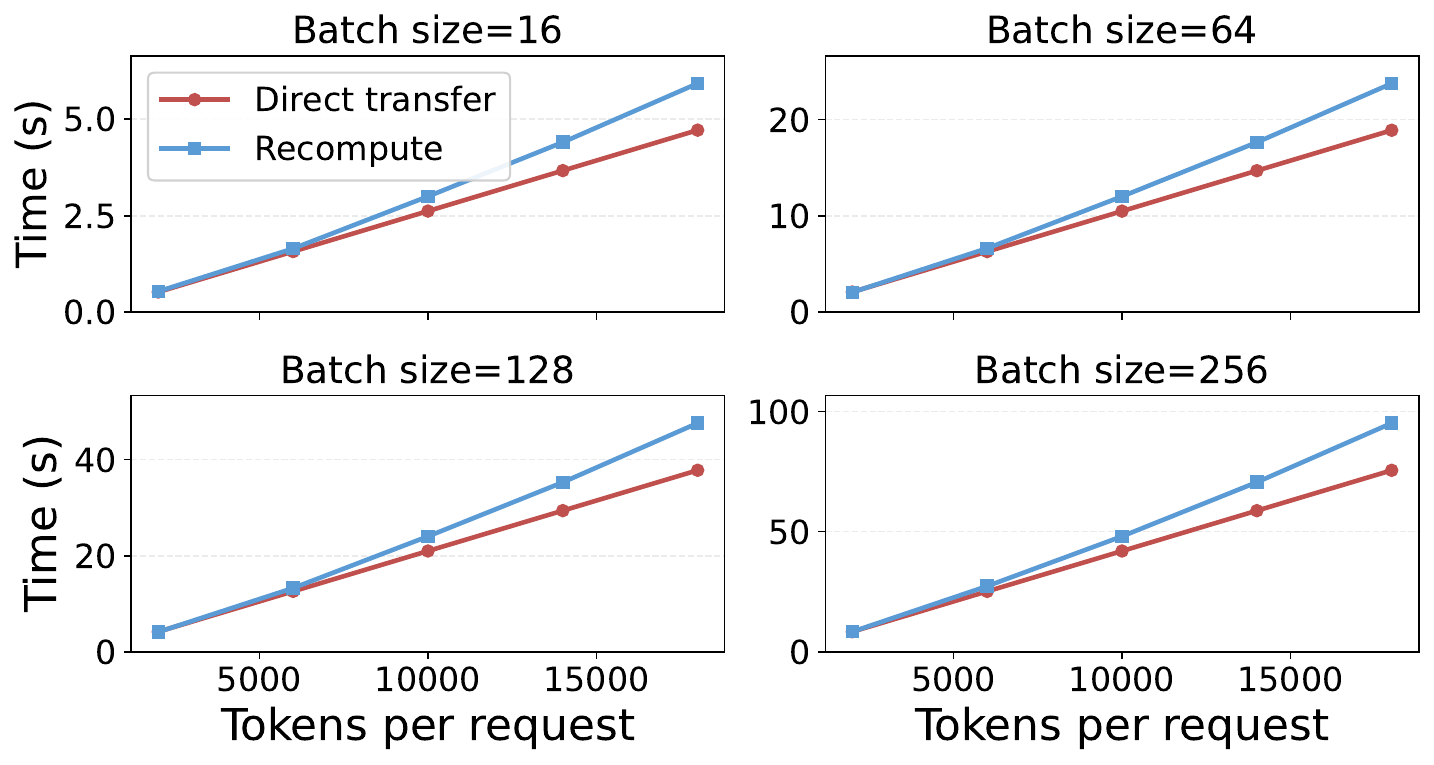}
    \caption{Time cost comparison between (i) direct KV cache transfer with one 200Gbps NIC and (ii) recomputation, across different batch sizes and generated-token lengths. TP=8.}
    \label{fig:kv-migration-cost}
\end{figure}

We present an example for KV-cache rebuilding overhead in Qwen-32B, including communication and synchronization for direct transfer, as well as forward-pass compute for recomputation, in Figure~\ref{fig:kv-migration-cost}. In general, direct transfer is more efficient for larger KV cache sizes (e.g., more generated tokens and/or larger batch sizes), while recomputation becomes favorable when the cache is smaller or communication bandwidth is a limiting factor. It is worth noting that this example assumes a single NIC with an aggregated bandwidth of 200 Gbps for KV-state transfer. In practical scenarios, where the layout between instances allows for multiple NICs, direct transfer can achieve significantly higher speeds.

Using profiled cost curves under various settings, our runtime dynamically selects the lower-cost option for each reconfiguration event.

\section{OCS Reconfiguration Time}\label{app:ocs_reconfig_time}

\begin{table}[t]
\centering
\footnotesize
\resizebox{0.85\columnwidth}{!}{
\begin{tabular}{lcc}
\hline
\textbf{OCS Type} & \textbf{Reconfig. delay (ms)} & \textbf{Radix ports} \\ \hline
RotorNet (InFocus)~\cite{mellette2020expanding} & 0.01 & 128 \\
3D MEMS (Calient)~\cite{ryf20011296mems} & 10 & 320 \\
Piezo (Polatis)~\cite{POLATIS_7000} & 25 & 576 \\
Liquid crystal (Coherent)~\cite{coherent2024ocs,ding2025photonic} & 100 & 512 \\
Robotic (Telescent)~\cite{wang2023topoopt} & 120000 & 1008 \\ \hline
\end{tabular}%
}
\caption{Representative reconfiguration delays and radix port counts for commodity OCS technologies.}
\label{tab:ocs_reconfig_time}
\end{table}

Table~\ref{tab:ocs_reconfig_time} summarizes the reconfiguration delays and radix port counts of various commodity OCS technologies.

\section{Model Configuration Summary}\label{app:model_config}

Table~\ref{tab:qwen25_configs} summarizes the key architectural hyperparameters of the Qwen variants used in this paper.
We report the training-time parallelism configurations used in our experiments.
For Qwen-72B training, we primarily use TP=8, PP=2, and CP=3, and scale DP accordingly.
For Qwen-57B-A14B training, we primarily use TP=4 and EP=16, and scale DP accordingly.

\begin{table}[t]
\resizebox{\columnwidth}{!}{%
\begin{tabular}{lcccccc}
\hline
\textbf{Model} & \textbf{Total} & $\mathbf{d_{model}}$ & \textbf{Layers} & \textbf{FFN intermediate} & \textbf{\# Query heads} & \textbf{\# KV heads} \\ \hline
Qwen2.5-14B & 14.7B & 5120 & 48 & 13824 & 40 & 8 \\
Qwen2.5-32B & 32.5B & 5120 & 64 & 27648 & 40 & 8 \\
Qwen2-57B-A14B (MoE) & 57.41B & 3584 & 28 & $2560 \times 64$ & 28 & 4 \\
Qwen2.5-72B & 72.7B & 8192 & 80 & 29568 & 64 & 8 \\
Qwen3-235B-A22B (MoE) & 235B & 4096 & 94 & $1536 \times 128$ & 64 & 4 \\
\hline
\end{tabular}%
}
\caption{Model configuration summary. For MoE, the FFN intermediate is written as (expert dim) $\times$ (num experts).}
\label{tab:qwen25_configs}
\end{table}

\section{OCS Port Allocation}
\label{app:portalloc}

This appendix describes \sys's OCS port/circuit allocation. We present (i) a tier-agnostic allocation abstraction
shared by OCS-Agg and OCS-Core, and (ii) a bounded-time greedy allocator for circuit allocation.

\noindent\textbf{Agg/Core relationship.}
In each period, \sys produces both an OCS-Agg plan (within each PoD, including ToR--uplink lift/drop) and an OCS-Core plan
(across PoDs). The two tiers are coupled: inter-PoD Core capacity incident to a PoD must be supportable by that PoD's
available boundary uplinks and the corresponding lift/drop access capacity. We therefore accept and commit a plan only
if the resulting Agg/Core allocations are jointly feasible under the current fabric state.

\subsection{Problem Abstraction}
\label{app:portalloc-abstraction}

For each upcoming communication period, \sys produces a sparse directed demand matrix
\(\{D_{u\rightarrow v}\}_{u,v\in V, u\neq v}\), where \(D_{u\rightarrow v}\) is the predicted bytes sent from endpoint
\(u\) to \(v\) during the period.
For OCS-Core, endpoints are PoDs.
For OCS-Agg, we use a ToR-centric view: each PoD is solved independently, and \(V\) includes the PoD's ToRs as well as
PoD-boundary uplink endpoints (i.e., the attachment points to OCS-Core).
This abstraction covers both intra-PoD ToR--ToR demands and inter-PoD access demands (ToR--uplink) that lift/drop traffic
to/from OCS-Core.
In practice, demands are represented sparsely as a directed edge set \(E_{\rightarrow}\subseteq V\times V\)
(e.g., after \textsc{AggregateDemand}); for \((u,v)\notin E_{\rightarrow}\), we treat \(D_{u\rightarrow v}=0\).
(When constructing undirected candidate pairs, the implementation enumerates from the sparse demanded edges and their
reverses, rather than all \(|V|^2\) pairs.)

A circuit is a discrete capacity unit between an endpoint pair. Each circuit is full duplex and provides line rate
\(B_{\text{link}}\) in each direction.
Let \(\deg(v)\) denote the number of allocatable OCS ports at endpoint \(v\) after applying port reservations.

\noindent\textbf{Decision variables.}
We allocate integer circuit multiplicities \(c_{uv}\ge 0\) over endpoint pairs, where \(c_{uv}=c_{vu}\) denotes the
number of parallel full-duplex circuits between endpoints \(u\) and \(v\).

\noindent\textbf{Port budget constraint.}
Each circuit consumes one port at each endpoint, hence:
\begin{align}
\sum_{u: u\neq v} c_{uv} \le \deg(v) \qquad \forall v\in V. 
\label{eq:port-constraint}
\end{align}

\noindent\textbf{Template scope.}
The selected topology template \(tpl\) determines which pairs are OCS-eligible via \(\textsc{Eligible}(tpl,u,v)\)
(and may additionally impose per-pair or role constraints). This appendix focuses on allocating \(c_{uv}\) on eligible
pairs under the port budget (i.e., we implicitly enforce \(c_{uv}=0\) when \(\neg \textsc{Eligible}(tpl,u,v)\)).

\subsection{Greedy Circuit Allocation}
\label{app:portalloc-heuristic}

Exact optimization over integer circuit multiplicities is expensive at scale, so \sys uses a greedy allocator. The
algorithm allocates circuits one-by-one until no demanded eligible pair has free ports at both endpoints. The number of
iterations is bounded by \(\frac{1}{2}\sum_{v\in V}\deg(v)\); moreover, the demand graph is usually sparse after
\textsc{AggregateDemand}.

Given directed demands but full-duplex circuits, we score bottlenecks at the endpoint-pair granularity. For each eligible
pair \(\{u,v\}\), define a completion-time proxy:
{\footnotesize
\begin{align}
T^{\text{pair}}[u,v]=
\begin{cases}
\max\!\left(\dfrac{D_{u\rightarrow v}}{c_{uv}B_{\text{link}}},\;
            \dfrac{D_{v\rightarrow u}}{c_{uv}B_{\text{link}}}\right),
& c_{uv}>0,\\[6pt]
+\infty, & c_{uv}=0 \land (D_{u\rightarrow v}+D_{v\rightarrow u}>0),\\
0,       & c_{uv}=0 \land (D_{u\rightarrow v}+D_{v\rightarrow u}=0).
\end{cases}
\label{eq:pair-completion-proxy}
\end{align}
}
Intuitively, \(T^{\text{pair}}\) approximates the time to serve the slower direction on the pair using the currently
allocated parallel circuits; a demanded pair with zero circuits is treated as an infinite bottleneck.

At each step, \textsc{AllocateCircuits} (Algorithm~\ref{alg:allocatecircuits}) selects the demanded eligible pair with the largest \(T^{\text{pair}}\) among
those with at least one free port on both endpoints, and allocates one additional circuit to that pair. When multiple
pairs have \(T^{\text{pair}}=+\infty\), ties are broken by the larger directional demand
\(\max(D_{u\rightarrow v},D_{v\rightarrow u})\).
For certain templates (e.g., bipartite structures), the implementation may optionally add a utilization-aware tie-breaker
to reduce port monopolization by hot endpoints and preserve feasible matchings.

After allocation, the system checks hard feasibility constraints (e.g., device/template constraints and that the planned
ports remain free at commit time) before committing the plan. It also enforces Core--Agg consistency for inter-PoD
traffic: the planned OCS-Core capacity must be supportable by OCS-Agg ToR--uplink access capacity in the corresponding
source and destination PoDs.

\begin{algorithm}[t]
\caption{\textsc{AllocateCircuits} (pair bottleneck greedy on OCS-eligible pairs)}
\footnotesize
\label{alg:allocatecircuits}
\begin{algorithmic}[1]
\Procedure{AllocateCircuits}{$tpl, G, S$}
\State $(V, E_{\rightarrow}, D_{\rightarrow}) \gets G$
\State $(\deg(\cdot), B_{\text{link}}) \gets S$
\State $C \gets \mathbf{0}$ \Comment{symmetric (full-duplex) circuit multiplicities}
\ForAll{$v\in V$} \State $p[v] \gets \deg(v)$ \EndFor

\While{true}
    \State $\mathcal{P} \gets \{\{u,v\}\mid u\neq v \land \textsc{Eligible}(tpl,u,v) \land p[u]>0 \land p[v]>0 \land (D_{\rightarrow}(u,v)+D_{\rightarrow}(v,u)>0)\}$
    \If{$\mathcal{P}=\emptyset$} \State \textbf{break} \EndIf

    \Statex \Comment{Pick pair with max \textsc{PairTime}; break ties by \textsc{MaxDirDemand}.}
  \State $\{u^\star,v^\star\} \gets
\arg\max_{\{u,v\}\in \mathcal{P}}
\begin{aligned}[t]
\Big(&\textsc{PairTime}(u,v,C,D_{\rightarrow},B_{\text{link}}),\\
     &\textsc{MaxDirDemand}(u,v,D_{\rightarrow})\Big)
\end{aligned}$

    \State $a \gets \min(u^\star,v^\star)$; \ $b \gets \max(u^\star,v^\star)$
    \State $C[a][b] \gets C[a][b] + 1$; \ $C[b][a] \gets C[b][a] + 1$
    \State $p[a] \gets p[a] - 1$; \ $p[b] \gets p[b] - 1$
\EndWhile
\State \Return $C$
\EndProcedure

\Function{PairTime}{$u,v,C,D_{\rightarrow},B_{\text{link}}$}
    \State $c \gets C[u][v]$
    \State $d_{uv} \gets D_{\rightarrow}(u,v)$; \ $d_{vu} \gets D_{\rightarrow}(v,u)$
    \If{$d_{uv}+d_{vu}=0$}
        \State \Return $0$
    \ElsIf{$c=0$}
        \State \Return $+\infty$
    \Else
        \State \Return $\max(d_{uv}, d_{vu})/(c\cdot B_{\text{link}})$
    \EndIf
\EndFunction

\Function{MaxDirDemand}{$u,v,D_{\rightarrow}$}
    \State \Return $\max(D_{\rightarrow}(u,v), D_{\rightarrow}(v,u))$
\EndFunction
\end{algorithmic}
\end{algorithm}

\section{Tree-Style Broadcast for Weight Synchronization}
\label{app:weightsync-tree}

This appendix describes \fabric's two-level, circuit-scheduled broadcast for disseminating weight shards.
Each shard is disseminated via its own tree; over many shards, the system materializes a temporary dissemination
\emph{forest}, and tears it down after synchronization completes.

Weight synchronization disseminates a large parameter shard from \train endpoints that hold the shard to all \gen
endpoints that require it. A naive one-to-all design (each \train sender directly to every \gen receiver) scales poorly
because inter-PoD transfers must traverse the PoD boundary, where Core-facing OCS ports are limited and can easily become
hotspots. \fabric therefore materializes a temporary \emph{two-level dissemination overlay} with ample scheduling slack:
Phase~A seeds a small set of \gen receivers across PoDs using OCS-Core; Phase~B completes intra-PoD fanout inside each
destination PoD.

\noindent\textbf{Endpoints and PoD-boundary concurrency budgets.}
Endpoints are GPUs/server NICs (ranks); let \(\textsc{pod}(x)\) denote the PoD containing endpoint \(x\).
OCS-Core constraints are enforced at the \emph{PoD boundary}: each PoD \(p\) has a budget \(d_p\) that caps the number of
\emph{concurrently active inter-PoD circuits} incident to \(p\) (counting one unit at both the source and destination
PoDs). Equivalently, \(d_p\) captures the scarce boundary-facing OCS-Core port capacity available for simultaneous use.
Gateway endpoints (when used) only affect which endpoints terminate boundary-facing circuits; budgets are still accounted
per PoD. In addition, realizing a Phase~A inter-PoD OCS-Core transfer may require intra-PoD connectivity to reach the boundary-facing uplinks (e.g., via OCS-Agg circuits between ToRs and
uplinks), in both the \train source PoD and the \gen destination PoD.

\subsection{Phase A: Cross-PoD Seeding on OCS-Core}
\label{app:weightsync-crosspod}

\noindent\textbf{Senders, destinations, and seeds.}
Let \(\mathcal{S}\) be the set of \train-side sender candidates that hold the shard (e.g., shard owners or replicas).
Let \(\mathcal{D}\) be the set of \gen endpoints that ultimately need the shard.
Phase~A delivers the shard only to a small set of \emph{seed receivers} \(\mathcal{R}\subseteq \mathcal{D}\) (typically a
constant number per destination PoD); Phase~B then fans out from seeds to all remaining endpoints within each destination
PoD, avoiding redundant cross-PoD replication.

\noindent\textbf{Overlay construction and budget-aware scheduling.}
Phase~A constructs an inter-PoD overlay on endpoints, in which each scheduled edge \((u,v)\) satisfies
\(\textsc{pod}(u)\neq \textsc{pod}(v)\).
We activate a subset of roots \(\mathcal{S}_{\text{act}}\subseteq \mathcal{S}\) and assign each seed to exactly one
active root: \(\mathcal{R}=\bigsqcup_{s\in\mathcal{S}_{\text{act}}}\mathcal{R}_s\).
Because PoD-boundary budgets bound \emph{concurrency}, Phase~A may run in multiple \emph{waves}. Let \(\mathcal{E}_t\) be
the set of inter-PoD edges active in wave \(t\), and let \(c_{uv}^{(t)}\ge 1\) be the number of circuits allocated to
\((u,v)\in\mathcal{E}_t\).
Feasibility requires, for every PoD \(p\) and wave \(t\),
\[
\sum_{(u,v)\in\mathcal{E}_t:\ \textsc{pod}(u)=p} c_{uv}^{(t)} \;+\;
\sum_{(u,v)\in\mathcal{E}_t:\ \textsc{pod}(v)=p} c_{uv}^{(t)}
\;\le\; d_p.
\]

\noindent\textbf{Circuit allocation and completion time.}
Each scheduled inter-PoD edge \((u,v)\) is realized by \(c_{uv}^{(t)}\) parallel point-to-point circuits of line rate
\(B_{\text{link}}\). Let \(S_{uv}^{(t)}\) be the data volume sent on \((u,v)\) in wave \(t\) (e.g., shard chunks).
Assuming circuit-level isolation and sufficiently large chunks, the wave completion time is well-approximated by the
slowest edge:
\[
T_{\text{A}}^{(t)} \approx \max_{(u,v)\in\mathcal{E}_t} \frac{S_{uv}^{(t)}}{c_{uv}^{(t)}\cdot B_{\text{link}}},\qquad
T_{\text{A}} \approx \sum_t T_{\text{A}}^{(t)}.
\]
Given a fixed eligible edge set (the Phase~A overlay), \fabric allocates at least one circuit per scheduled edge and
greedily adds circuits to equalize per-edge bottlenecks while respecting the per-PoD incidence budgets \(\{d_p\}\)
(similar in spirit to the bottleneck-equalization heuristic in Appendix~\ref{app:portalloc}).

\subsection{Phase B: Intra-PoD Fanout on OCS-Agg}
\label{app:weightsync-intrapod}

Once a seed in a destination PoD receives the shard, it disseminates it to the remaining \gen ranks within the same PoD.
To accelerate cross-rack fanout and avoid bottlenecks at a single root NIC, rack uplink, or ToR, \fabric may materialize
intra-PoD OCS-Agg circuits (or use rack-local EPS when sufficient). This allows Phase~A to reduce cross-PoD boundary
occupancy by seeding only \(\mathcal{R}\).
Within each destination PoD, \fabric schedules an intra-PoD dissemination tree---or multiple trees rooted at different
seeds---so that each seed serves only a fraction of receivers.

\section{Optimization Solving Time}\label{app:optim_solve_time}
\begin{figure}[t]
  \centering
  \includegraphics[width=0.235\textwidth]{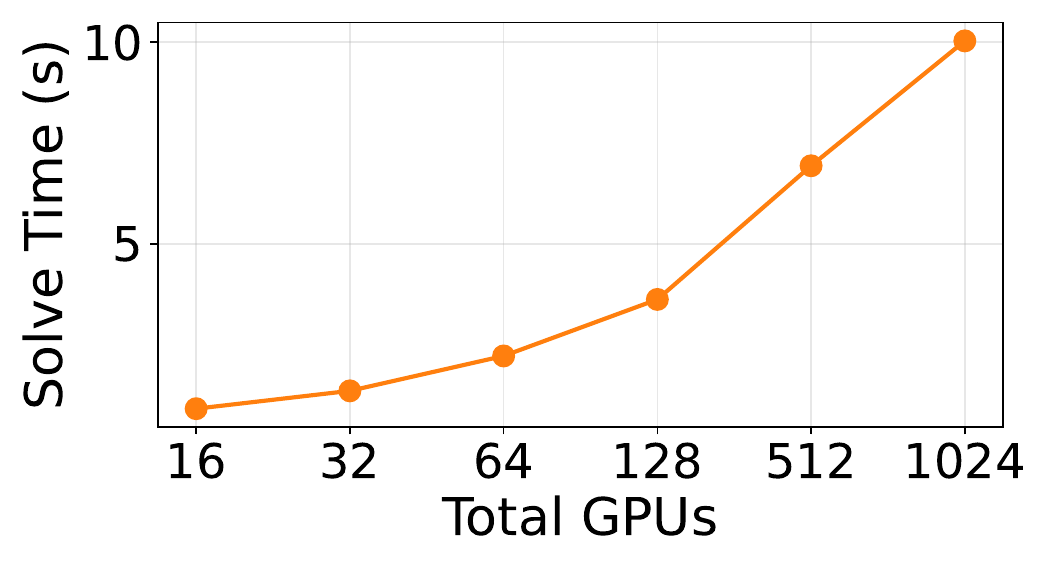}
  \vspace{-0.08in}
  \caption{Optimization solving time across GPU scales. The total number of requests is set to \(16 \times\) \#GPUs.}
  \label{fig:solve_time}
\end{figure}

Figure~\ref{fig:solve_time} reports the solving time of the optimization problem in our compute scheduler across different GPU scales for Qwen-57B-MoE model. Even at a scale of 1024 GPUs with a large request volume, the solver finishes in under 11\,s, which can be largely overlapped with phases such as weight synchronization. Moreover, during a generation step, the number of active requests naturally decreases as requests complete; thus, for the same GPU scale, the optimization solving time typically decreases accordingly.

\section{Network Component Price}\label{app:network_price}
Table~\ref{tab:cost_components} summarizes indicative prices for different network components, compiled from a search of recent market listings and publicly available pricing information. These values are used as baseline inputs for our cost comparison and do not represent vendor quotes.

\begin{table}[t]
\centering
\resizebox{\columnwidth}{!}{%
\begin{tabular}{lcccccc}
\toprule
\textbf{\begin{tabular}[c]{@{}l@{}}Link\\ Bandwidth\end{tabular}} & \textbf{\begin{tabular}[c]{@{}c@{}}Trans-\\ ceiver (\$)\end{tabular}} & \textbf{NIC (\$)} & \textbf{\begin{tabular}[c]{@{}c@{}}Elec.\\ Switch\\ Port (\$)\end{tabular}} & \textbf{\begin{tabular}[c]{@{}c@{}}OCS\\ Port\\ (\$)\end{tabular}} & \textbf{\begin{tabular}[c]{@{}c@{}}Patch\\ Panel\\ Port (\$)\end{tabular}} & \textbf{\begin{tabular}[c]{@{}c@{}}Fiber\\ Cable\\ (\$)\end{tabular}} \\ \midrule
100 Gbps & 249~\cite{fs_100g_cwdm4}  & 736~\cite{fs_cx5_100g}   & 215~\cite{ebay_edgecore_100g}  & 350~\cite{Diego2013comopt,calient_s320} & 100~\cite{wang2023topoopt} & 25~\cite{dell_cable_100g} \\
200 Gbps & 499~\cite{proficium_200g} & 1,291~\cite{newegg_cx6_200g}  & 876~\cite{nvidia_sn4600_200gb}  & 350~\cite{Diego2013comopt,calient_s320} & 100~\cite{wang2023topoopt} & 45~\cite{compufox_mpo12} \\
400 Gbps & 799~\cite{fs_400g_dr4}    & 1,710~\cite{dell_cx7_400g}  & 1,392~\cite{cisco_neux_64x400g}  & 350~\cite{Diego2013comopt,calient_s320} & 100~\cite{wang2023topoopt} & 65~\cite{fluxlight_mpo_400g} \\
800 Gbps & 1,599~\cite{fs_800g_dr8}   & 2,347~\cite{shi_cx8_800g}      & 1,731~\cite{Arista_switch_800g}    & 350~\cite{Diego2013comopt,calient_s320} & 100~\cite{wang2023topoopt} & 92~\cite{lightoptics_mpo16} \\ \bottomrule
\end{tabular}
}
\caption{Price of Network Components.}
\label{tab:cost_components}
\end{table}

\section{Failure Handling}\label{app:network_failure}

\begin{figure}[t]
  \centering
  \includegraphics[width=0.435\textwidth]{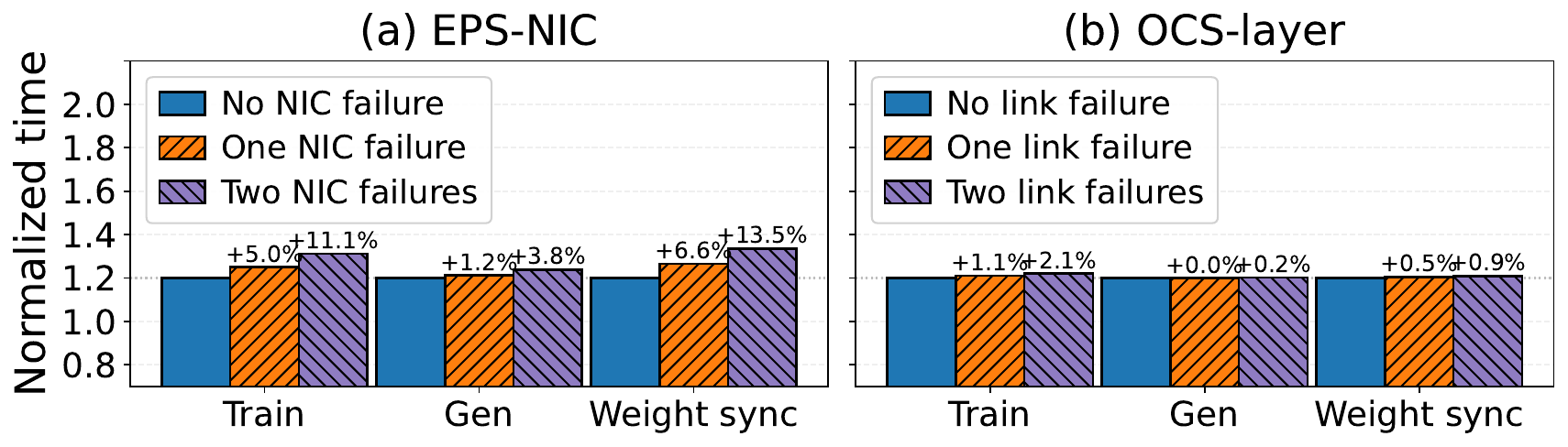}
  \vspace{-0.08in}
  \caption{Impact of link failures on end-to-end time (normalized) across RL stages for Qwen2-57B-MoE. We inject one or two failures on (a) NIC--EPS links and (b) OCS-layer links, and report the normalized \train, \gen, and \weightsync time.}
  \label{fig:failure}
\end{figure}

We evaluate \fabric's resiliency to link failures using Qwen2-57B-MoE across different RL stages, considering failures on the NIC--EPS link and on links in the OCS layer.
Figure~\ref{fig:failure} shows that a NIC--EPS link failure has a noticeable impact on communication-intensive stages: \train slows down by \(5.0\%\) (one failure) and \(11.1\%\) (two failures), while \weightsync degrades by \(6.6\%\) and \(13.5\%\), respectively.
This is because a NIC--EPS failure forces the affected GPU to reroute its traffic through the scale-up domain (e.g., NVLink/NVSwitch) to a healthy NIC, creating additional intra-node hops and oversubscribing the remaining NICs.

In contrast, \gen is less sensitive (only \(1.2\%\) and \(3.8\%\) slowdown) because it runs in AFD mode, where communication overhead can be largely overlapped with computation.

Failures in the OCS layer have a much smaller impact: \train increases by only \(1.1\%\) and \(2.1\%\) under one and two link failures, and \weightsync by \(0.5\%\) and \(0.9\%\), while \gen remains nearly unchanged (\(0.0\%\) and \(0.2\%\)).
This is because OCS-layer failures can often be masked by multi-path routing (e.g., ECMP), whereas a NIC--EPS failure directly reduces the available bandwidth at the endpoint and concentrates traffic on fewer NICs.

\section{Sensitivity Analysis}\label{app:sensitive}

\begin{figure}[t]
  \centering
\includegraphics[width=0.40\textwidth]{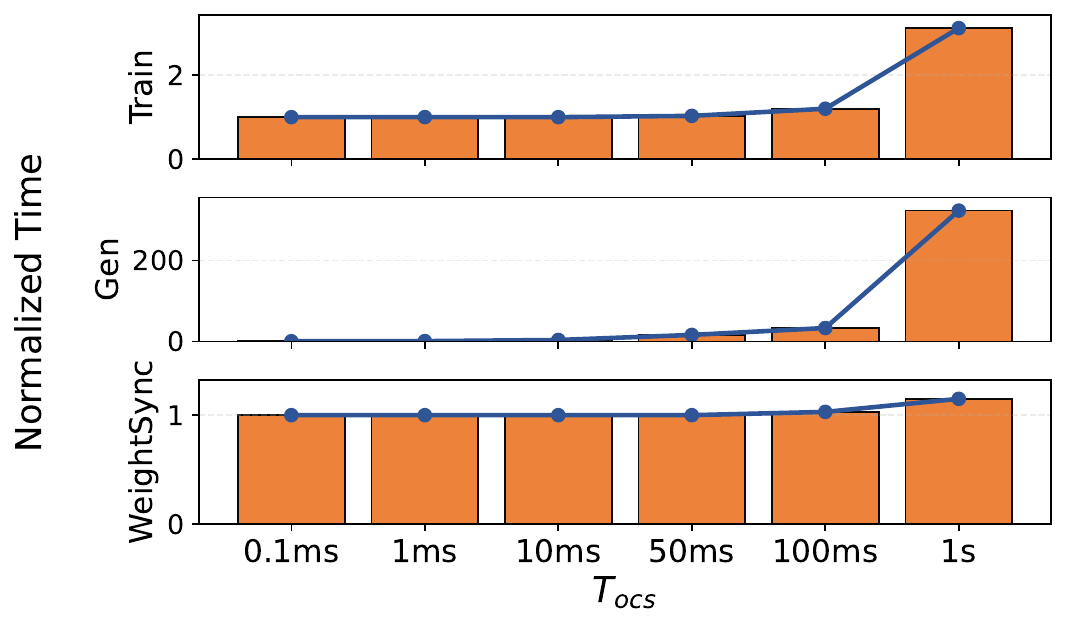}
  \vspace{-0.08in}
\caption{Sensitivity to OCS update time under fine-grained reconfiguration across RL stages.}
\label{fig:ocs_time_sensitivity} 
\end{figure}

\noindent\textbf{Sensitivity to \(T_{\text{ocs}}\).}
We sweep \(T_{\text{ocs}}\) over representative values and report the normalized stage time of \train, \gen, and \weightsync for Qwen-235B-A22B (Figure~\ref{fig:ocs_time_sensitivity}). We assume fine-grained reconfiguration in each stage---in contrast to our stage-dependent adaptive granularity---including in-execution updates when communication patterns shift (e.g., during AFD in \gen, MoE transitions from cross-server all-to-all to M2N back to attention, triggering a reconfiguration). Sensitivity is stage-dependent and matches the slack distributions in \S2. \train degrades gradually as \(T_{\text{ocs}}\) grows because reduced compute slack can no longer hide reconfiguration; it remains stable for \(T_{\text{ocs}} \le 100\,\mathrm{ms}\). \gen is highly sensitive: even millisecond-scale reconfiguration can become a bottleneck due to limited overlap slack, inflating latency by \(4\text{--}325\times\). \weightsync is minimally affected over practical ranges because it has ample slack and the completion time is nearly unchanged.

\begin{figure}[t]
  \centering
\includegraphics[width=0.445\textwidth]{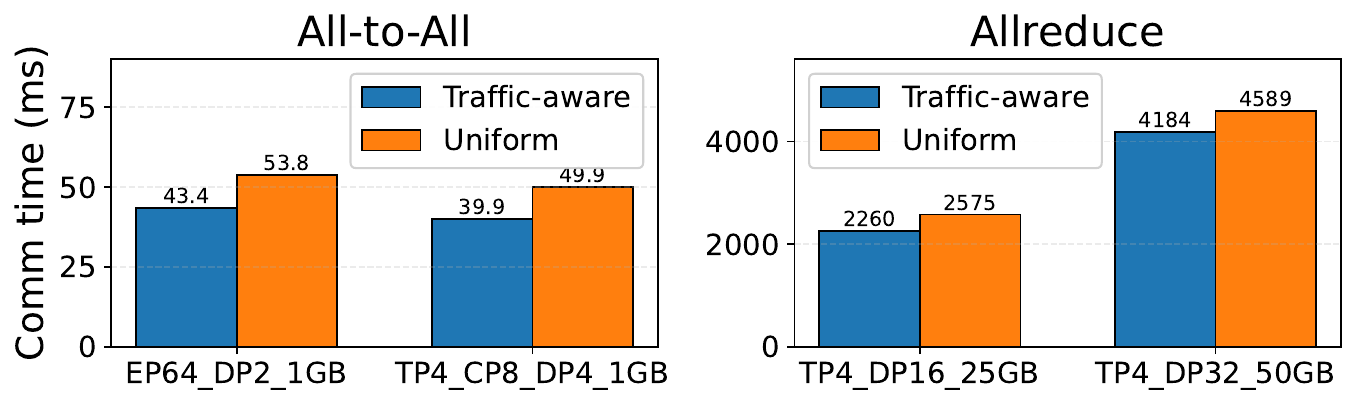}
  \vspace{-0.08in}
\caption{Collective performance of \fabric under traffic-aware allocation versus uniform allocation with 200\,Gbps links.
Labels denote the parallelism configuration, and sizes (1/25/50\,GB) are per-rank message sizes. All-to-all runs over the
EP/CP domain, while allreduce runs over the DP domain.}
\label{fig:alloc_policy_collectives} 
\end{figure}

\noindent\textbf{Traffic-aware allocation vs.\ uniform splitting.}
Figure~\ref{fig:alloc_policy_collectives} shows that \fabric's traffic-aware allocation outperforms uniform splitting
across the evaluated collectives and configurations. For all-to-all, traffic-aware reduces communication time by 19.3\%
on \texttt{EP64\_DP2\_1GB} and 20.0\% on \texttt{TP4\_CP8\_DP4\_1GB}. For allreduce, it reduces communication time by 12.1\%
on \texttt{TP4\_DP16\_25GB} and 8.8\% on \texttt{TP4\_DP32\_50GB}. Uniform splitting spreads scarce OCS ports (and thus circuit
capacity) evenly across endpoints, which can over-provision low-demand pairs while leaving hotspots under-provisioned. In
contrast, traffic-aware allocation concentrates circuits on high-demand pairs and dominant bottlenecks, improving
collective communication performance.

\end{document}